\documentclass[journal,10pt]{IEEEtran}
\usepackage{subfigure}
\PassOptionsToPackage{subfigure}{tocloft}
\usepackage{CJK}
\usepackage{algorithm}
\usepackage{algorithmic}
\usepackage{amsmath}
\usepackage{amssymb}
\usepackage{graphicx}
\usepackage{epstopdf}
\usepackage{multirow}
\usepackage{stfloats}
\usepackage{cite}
\usepackage{color}
\usepackage[normalem]{ulem}
\usepackage{enumerate}
\usepackage{bbm}
\usepackage{times,amsmath,amsthm,color,amssymb,graphicx,epsfig,algorithm,algorithmic,bm,multirow,cite}
\theoremstyle{definition}

\floatname{algorithm}{Algorithm}

\title{Power-Aware Sparse Reflect Beamforming in Active RIS-aided Interference Channels}

\author{Ruizhe Long, Hu Zhou and Ying-Chang Liang, \emph{Fellow, IEEE}
\thanks{

This work has been submitted to the IEEE for possible publication. Copyright may be transferred without notice, after which this version may no longer be accessible. \emph{(Corresponding author: Ying-Chang Liang)}

R.~Long and H. Zhou are with the National Key Laboratory of Science and Technology on Communications and also the Center for Intelligent Networking and Communications (CINC), University of Electronic Science and Technology of China (UESTC), Chengdu 611731, China (e-mail: ruizhelong@gmail.com, huzhou@std.uestc.edu.cn).

Y.-C. Liang is with the Center for Intelligent Networking and Communications (CINC), University of Electronic Science and Technology of China (UESTC), Chengdu 611731, China (e-mail: liangyc@ieee.org).}
}

\begin{document}
\maketitle
\begin{abstract}
  Active reconfigurable intelligent surface (RIS) has attracted significant attention in wireless communications, due to its reflecting elements (REs) capable of reflecting incident signals with not only phase shifts but also amplitude amplifications. In this paper, we are interested in active RIS-aided interference channels in which $K$ user pairs share the same time and frequency resources with the aid of active RIS. Thanks to the promising amplitude amplification capability, activating a moderate number of REs, rather than all of them, is sufficient for the active RIS to mitigate cross-channel interferences. Motivated by this, we propose a power-aware sparse reflect beamforming design for the active RIS-aided interference channels, which allows the active RIS to flexibly adjust the number of activated REs for the sake of reducing hardware and power costs. Specifically, we establish the power consumption model in which only those activated REs consume the biasing and operation power that supports the amplitude amplification, yielding an $\ell_0$-norm power consumption function. Based on the proposed model, we investigate a sum-rate maximization problem and an active RIS power minimization problem by carefully designing the sparse reflect beamforming vector. To solve these problems, we first replace the nonconvex $\ell_0$-norm function with an iterative reweighted $\ell_1$-norm function. Then, fractional programming is used to solve the sum-rate maximization, while semidefinite programming together with the difference-of-convex algorithm (DCA) is used to solve the active RIS power minimization. Numerical results show that the proposed sparse designs can notably increase the sum rate of user pairs and decrease the power consumption of active RIS in interference channels.  
\end{abstract}

\section{Introduction}
\label{sec:intro}

Reconfigurable intelligent surface (RIS) has been envisioned as one of the enabling technologies for sixth-generation (6G) wireless communication~\cite{you2021towards}, due to its unique capability to reshape wireless channels and enhance capacity of wireless systems~\cite{Liang2019,DiRenzo2020}. More specifically, an RIS consists of  multiple reflecting elements (REs) that reflect incident signals with phase shifts and sometimes amplitude modifications, allowing to enhance the desired signals and to suppress the undesired signals. Consequently, RIS make it a success to promote spectrum and energy efficiency of wireless communications~\cite{Guo2020Weighted,Huang2019}.

Currently, based on whether its REs are able to reflect incident signals with amplitude amplification, RIS can be broadly classified into passive RIS~\cite{Guo2020Weighted,Huang2019} and active RIS~\cite{Long2021Active}.  In passive RIS, each RE primarily adjusts the phase shift of the incident signal via reflection, thereby facilitating favorable radio propagation by constructively (destructively) combining the desired (undesired) signals. However, passive RIS faces challenges in addressing the double fading attenuation problem in which the reflected signals undergo significant attenuation due to multiplicative fading loss in the cascaded channel~\cite{najafi2020physics}, limiting the service area of passive RIS. In active RIS, each RE preserves the capability of phase shifts and, meanwhile, can amplify the incident signal via additional reflection amplifier circuits~\cite{Wu2022Wideband,Rao2023An}. This allows active RIS to efficiently tackle double fading attenuation with the capability of amplitude amplification, and thus active RIS can significantly extend the RIS service area and boost the spectrum efficiency of users~\cite{Long2021Active,Zhang2023Active}. This emerging RIS technique has attracted considerable attention in academic research and found applications across various wireless communication scenarios. From enhancing the secrecy capacity of wireless channels~\cite{Dong2022Active} to improving power transfer in integrated wireless information and power systems~\cite{Gao2022Beamforming}, and even safeguarding primary users in cognitive radio networks from secondary user interference~\cite{Yang2023Active}, active RIS is reshaping wireless communication paradigms.

The importance of RIS is particularly pronounced with its applications in interference channels where  multiple users transmit their individual and independent information to the corresponding receivers with the same time and frequency resources. Addressing interference in wireless communications is a critical challenge, various techniques deployed at communication nodes have been proposed in the past decades, such as transmit beamfoming~\cite{Jorswieck2008Complete} and interference alignment~\cite{cadambe2008interference}. RIS offers a novel solution with reconfigurable radio propagation, since it directly destructs the cross channels that carry interference and significantly enhances interference management strategies. For example, with the assistance of passive RIS, the interference alignment technique can increase the number of interference-free transmissions from $\frac{K}{2}$~\cite{cadambe2008interference} to $K$~\cite{bafghi2022degrees} in a $K$-user interference channel. The study in~\cite{jiang2022interference} introduces an RIS-aided interference nulling strategy, demonstrating that a passive RIS using a reflect beamforming vector with $2K(K-1)$ REs can entirely mitigate interference when the direct channels between user pairs are blocked. Furthermore, a passive RIS interference mitigation design is proposed in~\cite{elmossallamy2021ris}, where the interference power minimization problem is addressed using the complex circle manifold.

Interference commonly arises in scenarios where multiple users are in close proximity, necessitating the effective management by RIS to handle the strong interference transmitted through high-gain cross channels between user pairs. Numerical results indicate that more than $2K(K-1)$ REs are needed for passive RIS to mitigate interference effectively when the cross channels are of high channel gains~\cite{elmossallamy2021ris}. Regarding the double fading attenuation, the excessive interference makes it a challenging task to completely mitigate interference with passive RIS. Although numerous REs can be employed by passive RIS to mitigate the strong interference, it would lead to a large size and an increase in the deployment costs. In this context, active RIS presents a more efficient solution for mitigating interference in densely populated networks. Thanks to its amplitude amplification capability, the active RIS usually requires fewer number of REs to completely mitigate interference than the passive one does. As validated in~\cite{jiang2022interference}, active RIS can achieve interference nulling with as few as $K(K-1)$ REs, showing the benefits of using active RIS in scenarios with challenging interference conditions. This efficiency not only highlights the superiority of active RIS in managing interference but also suggests a more efficient approach to deploy RIS in complex wireless environments.

However, integrating active RIS into interference channels comes with its own challenges, particularly concerning increased power consumption and hardware costs due to the additional requirements for reflection amplifier circuits~\cite{Wu2022Wideband,Rao2023An,Amato2018a}. To overcome these challenges, we propose a power-aware sparse reflect beamforming design that allows the active RIS to flexibly adjust the number of activated REs. In particular, we first investigate the capability of active RIS to mitigate interference under the maximum amplitude constraints on reflection coefficients, which reveals the active RIS with a higher maximum amplitude constraint can exploit less REs to completely mitigate the cross-channel interference. Subsequently, we establish a more practical power consumption model in which only the activated REs consume the additional power to amplify the incident signal, thereby reducing overall power consumption and hardware costs. This model leverages the sparsity of RE activation, enabling efficient determination of the necessary number of REs for achieving specific performance targets. Addressing the challenge of selecting the optimal REs for activation, we formulate two utility optimization problems: maximizing the sum rate of user pairs under a limited power budget and minimizing active RIS power consumption while meeting minimum rate requirements. 

This paper presents a novel study on active RIS-aided interference channels, incorporating a modified power consumption model for active RIS. To the best of our knowledge, this is the first work to consider the sparsity in the power consumption of active RIS in interference channels. The main contributions of this paper are summarized as follows

\begin{itemize}

  \item First, we propose a modified power consumption model for the active RIS, where the active RIS is allowed to close parts of inefficient REs for the sake of energy saving, and only those activated REs consume additional biasing and circuit operation power. This model later leads to two kinds of power-aware designs on active RIS.
    
  \item Second, we propose a power-aware sparse reflect beamforming design for the purpose of maximizing the sum-rate of user pairs in interference channels under the limited power budget and the maximum amplitude constraints. To tackle the nonconvexity caused by the $\ell_0$-norm constraint on the power consumption, an iterative reweighted $\ell_1$-norm method is first exploited to relax the $\ell_0$-norm constraint. Then, with the aid of fractional programming (FP), the sum-rate maximization can be efficiently solved.
  
  \item Third, we minimize the power consumption of the active RIS when all the user pairs meet the minimum rate requirements. The sparsity of the solution can also be recovered with the same $\ell_0$-norm relaxation. After the $\ell_0$-norm relaxation, the active power minimization problem can be solved with semidefinite programming (SDP) and the difference-of-convex algorithm (DCA).
  
  \item Finally, extensive simulation results are provided to compare the performance of the RIS-aided interference channels under various RIS setups. The results show that, by closing parts of REs, the proposed power-aware designs on active RISs can notably increase the sum rate of user transmissions even under a stringent power budget, and can also dramatically reduce the power consumption of active RISs in interference channels.
\end{itemize}

The rest of this paper is organized as follows. In Section~\ref{sec:SystemModel}, we introduce the system model of RIS-aided interference channels. In Section~\ref{sec:Nulling}, we investigate the interference nulling problem for active and passive RIS. In Section~\ref{sec:powermodel}, we propose the modified power consumption model, and consider two kinds of power-ware designs of active RIS in interference channels. In Section~\ref{sec:NormRelaxation}, we introduce the $\ell_0$-norm relaxation method. In Section~\ref{sec:Algs}, the complete algorithms for the sum-rate maximization and the active RIS power minimization are given. In Section~\ref{sec:comparison} and Section \ref{sec:sim}, comparisons and numerical results between various RIS designs are presented, respectively. Finally, the paper is concluded in Section \ref{sec:con}.



The major notations in this paper are listed as follows: Lowercase, boldface lowercase, and boldface uppercase letters, such as  $x$, $\mathbf{x}$, and $\mathbf{X}$, denote scalars, vectors, and matrices, respectively. $|x|$ denotes the absolute value of $x$, and $|\mathbf{x}|$ denotes the vector element-wise absolute value of $\mathbf{x}$. $\|\mathbf{x}\|$ denotes the norm of vector $\mathbf{x}$. ${\cal{CN}}(\mu, \sigma^2)$ denotes the complex Gaussian distribution with mean $\mu$ and variance $\sigma^2$. $\mathbb{E}[\cdot]$ denotes the statistical expectation. $x^{\ast}$ denotes the conjugate of $x$. $\mathbf{X}^{\mathrm{T}}$ and $\mathbf{X}^{\mathrm{H}}$ denotes the transpose and conjugate transpose of matrix $\mathbf{X}$, respectively. $\mathrm{diag}(\mathbf{x})$ returns a diagonal matrix that puts $\mathbf{x}$ on the main diagonal. $\mathbf{B}=\mathrm{blkdiag}(\mathbf{X}_1,...,\mathbf{X}_N)$ returns the block diagonal matrix created by aligning the input matrices $\mathbf{X}_1,...,\mathbf{X}_N$ along the diagonal of $\mathbf{B}$.

\section{System Model}\label{sec:SystemModel}
\begin{figure}[t]
  \centering 
  \includegraphics[width=.99\columnwidth]{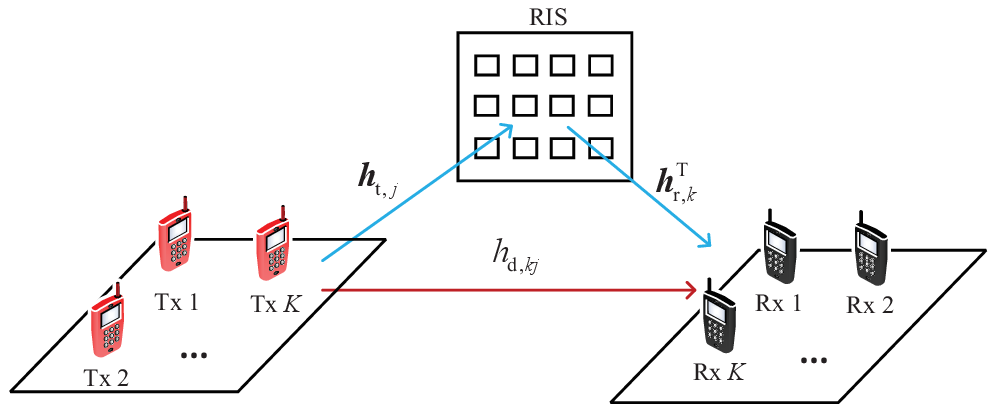}
  \caption{RIS-aided interference channels}
  \label{fig:system_model}
\end{figure}

This paper considers RIS-aided $K$-user pairs interference channels, where the RIS equipped with $Q$ REs is designed to help $K$ single-antenna transmitter (Tx) and receiver (Rx) pairs communicate at the same time over a common frequency band. As illustrated in Fig.~\ref{fig:system_model}, the channel from Tx $j$ to Rx $k$ is denoted by ${h}_{\mathrm{d},kj}$, while the forward channel from Tx $j$ to the RIS and the backward channel from the RIS to Rx $k$ are denoted by $\boldsymbol{h}_{\mathrm{t},j}\in\mathbb{C}^{Q \times 1}$ and $\boldsymbol{h}_{\mathrm{r},k}\in\mathbb{C}^{Q \times 1}$, respectively. We assume that a centralized controller collects all channel state information (CSI) with the channel training~\cite{chen2019channel}. In order to investigate the system performance that can be achieved by optimizing the RIS, we assume the CSI is perfectly known at the centralized controller. 

The reflection coefficients of the RIS constitute a diagonal matrix $\boldsymbol{\Phi} = \mathrm{diag}[\alpha_1 e^{j\theta_1},\cdots,\alpha_Q e^{j\theta_Q}]$ with $\alpha_q$ and $\theta_q$ being the amplitude and the phase of the reflection coefficient for the $q$-th RE. If the RIS is an active one that is able to amplify incident signals with the aid of biasing power sources, the amplitude of the reflection coefficients can be greater than unity, i.e., $\alpha_q \in [0,\alpha_{\max}],\alpha_{\max}\geq 1$, $\forall q$, but it is limited by the maximum amplitude $\alpha_{\max}$~\cite{Long2021Active}. In addition, due to the biasing power sources, the active RIS inevitably introduces thermal noises at the RIS side. If the RIS is a passive one that is not able to amplify the incident signal, the amplitude of the reflection coefficients is thus $\alpha_q \in[0,1],\forall q$. This paper mainly investigates the system performance with the assistance of active RIS.

In interference channels, Tx $k$ aims to send a transmit symbol $s_k$ to Rx $k$ in the presence of transmissions from other user pairs that share the same time and frequency resources. By means of the active RIS, the received signal $y_{\mathrm{a},k}$ at each Rx $k$ is thus given by
\begin{align}
  y_{\mathrm{a},k} & = \left(h_{\mathrm{d},kk}+\boldsymbol{h}_{\mathrm{r},k}^{\mathrm{T}}\boldsymbol{\Phi}\boldsymbol{h}_{\mathrm{t},k}\right)s_k + \sum_{j\neq k}^{K}\left(h_{\mathrm{d},kj} +   \boldsymbol{h}_{\mathrm{r},k}^{\mathrm{T}}\boldsymbol{\Phi}\boldsymbol{h}_{\mathrm{t},j}\right)s_j\nonumber \\
  & \qquad \qquad\qquad\qquad\qquad\quad   + \boldsymbol{h}_{\mathrm{r},k}^{\mathrm{T}}\boldsymbol{\Phi}\boldsymbol{n}_\mathrm{r} +n_k, \nonumber\\
  & = \left(h_{\mathrm{d},kk} + \boldsymbol{h}^{\mathrm{H}}_{\mathrm{b},kk}\boldsymbol{a} \right)s_k + \sum_{j\neq k}^{K}\left(h_{\mathrm{d},kj} + \boldsymbol{h}^{\mathrm{H}}_{\mathrm{b},kj}\boldsymbol{a} \right)s_j \nonumber\\
  & \qquad \qquad \qquad\qquad\qquad\quad +  \boldsymbol{n}_{\mathrm{r}}^\mathrm{T}\mathrm{diag}(\boldsymbol{h}_{\mathrm{r},k})\boldsymbol{a} + n_k, \label{eq:y_k}
\end{align}
where $\boldsymbol{n}_\mathrm{r}\sim\mathcal{CN}(\mathbf{0},\sigma^{2}_{r}\mathbf{I}_{Q})$ is the additive white Gaussian noise (AWGN) at the active RIS, $n_k\sim\mathcal{CN}(0,\sigma^{2}_{s})$ is the AWGN at Rx~$k$, $\boldsymbol{a} = \mathrm{diag}(\boldsymbol{\Phi})$ is the
reflect beamforming vector with the reflection coefficients being listed in the vector form, $\boldsymbol{h}_{\mathrm{b},kj}= \mathrm{diag}(\boldsymbol{h}_{\mathrm{t},j}^{*})\boldsymbol{h}_{\mathrm{r},j}^{*}$ is the effective cascaded channel, and $\boldsymbol{h}^{\mathrm{H}}_{\mathrm{b},kj}\boldsymbol{a}$ is the RIS-aided channel from Tx $j$ to Rx $k$. The transmit power of each Tx $k$ is denoted by $p_k$, i.e., $\mathbb{E}\left[|s_k|^2\right]=p_k$. The achievable rate of the $k$-th pair is thus given by
\begin{align}
R_{\mathrm{a},k} \!=\! \log\left(\!1 \!+ \!\frac{p_k\left|h_{\mathrm{d},kk} \!+\! \boldsymbol{h}^{\mathrm{H}}_{\mathrm{b},kk}\boldsymbol{a}\right|^2}{\sum\limits_{j\neq k}^{K}p_j\left| h_{\mathrm{d},kj} \!+\! \boldsymbol{h}^{\mathrm{H}}_{\mathrm{b},kj}\boldsymbol{a}\right|^2\!+\!\sigma^{2}_{r}\boldsymbol{a}^\mathrm{H}\boldsymbol{R}_{\mathrm{r},k}\boldsymbol{a}\!+\!\sigma^{2}_{s}}\!\right), \nonumber
\end{align}
where $\boldsymbol{R}_{\mathrm{r},k}$ is a diagonal matrix with $\boldsymbol{R}_{\mathrm{r},k} = \mathrm{diag}(|h_{\mathrm{r},1}|^2,...,|h_{\mathrm{r},K}|^2)$.
For the case that the RIS is a passive one, no thermal noise is introduced at the RIS, and thus $\sigma_r^2 = 0$. 

With the passive RIS, the achievable rate of the $k$-th pair is thus given by 
\begin{align}\label{eq:R_k_passive}
  R_{\mathrm{p},k} = \log\left(1+\frac{p_k\left|h_{\mathrm{d},kk} + \boldsymbol{h}^{\mathrm{H}}_{\mathrm{b},kk}\boldsymbol{a}\right|^2}{\sum\limits_{j\neq k}^{K}p_j\left|h_{\mathrm{d},kj} + \boldsymbol{h}^{\mathrm{H}}_{\mathrm{b},kj}\boldsymbol{a}\right|^2+\sigma^{2}_{s}}\right).
\end{align}

\section{RIS-aided Interference Nulling}\label{sec:Nulling}
As shown in the signal model~\eqref{eq:y_k}, $K$ pairs of signals go through the original channels as well as the RIS-aided channels, and RIS is able to tune the reflection coefficients of REs to mitigate the cross-channel interference between the Tx and Rx pairs. To do so, the interference channels between Tx~$j$ and Rx~$k$ should be null as follows
\begin{align}\label{eq:Nulling_cond}
  h_{\mathrm{d},kj} + \boldsymbol{h}^{\mathrm{H}}_{\mathrm{b},kj}\boldsymbol{a} = 0,\quad k = 1,\cdots,K,\quad \forall j\neq k. 
\end{align}
In other words, the reflect beamforming vector $\boldsymbol{a}$ should be designed to make the RIS-aided channels $\boldsymbol{h}^{\mathrm{H}}_{\mathrm{b},kj}\boldsymbol{a}$ and the original channels $h_{\mathrm{d},kj}$ have equal strength and opposite phase. By letting 
\begin{align}
  \boldsymbol{H}_\mathrm{b} &= [\boldsymbol{h}_{\mathrm{b},21},\cdots,\boldsymbol{h}_{\mathrm{b},K1},\cdots,\boldsymbol{h}_{\mathrm{b},1K},\cdots,\boldsymbol{h}_{\mathrm{b},(K-1)K}] ,\nonumber\\
  \boldsymbol{h}_\mathrm{d} &= [{h}_{\mathrm{d},21},\cdots,{h}_{\mathrm{d},K1},\cdots,{h}_{\mathrm{d},1K},\cdots,{h}_{\mathrm{d},(K-1)K}]^{\mathrm{T}},\nonumber
\end{align}
a compact form of~\eqref{eq:Nulling_cond} is given by 
\begin{align}\label{eq:Nulling_cond_compact}
  \boldsymbol{H}_\mathrm{b}^\mathrm{H} \boldsymbol{a} = - \boldsymbol{h}_\mathrm{d}.
\end{align}
We assume that the channel matrix $\boldsymbol{H}_\mathrm{b}$ is full column rank, i.e., $\mathrm{rank}(\boldsymbol{H}_\mathrm{b})=K(K-1)$. The assumption is reasonable when the RIS is equipped with a large scale of REs ($Q\geq K(K-1)$) and $\boldsymbol{h}^{\mathrm{H}}_{\mathrm{b},kj},~\forall k\neq j$ is linearly independent. Under this assumption, a feasible solution to $\boldsymbol{a}$ is obtained by solving the underdetermined linear system problem in~\eqref{eq:Nulling_cond_compact}, which is given by
\begin{align}\label{eq:a_cf}
  \boldsymbol{a} = -\boldsymbol{H}_\mathrm{b}\left(\boldsymbol{H}_\mathrm{b}^\mathrm{H}\boldsymbol{H}_\mathrm{b}\right)^{-1}\boldsymbol{h}_\mathrm{d}.  
\end{align}
This solution shows a necessary condition that the number of REs should be greater than the number of interference channels to achieve interference nulling via the RIS, i.e., $Q\geq K(K-1)$. Otherwise, the RIS has no sufficient degree of freedoms to completely mitigate the interference. When equipped with the insufficient number of REs, the RIS is still able to create interference-free channels with their DoF less than $K$ but greater than $K/2$~\cite{bafghi2022degrees}.

For the ideal case, the RIS is expected to mitigate all the interference among the user pairs, which can be achieved when the necessary condition is satisfied and no other constraints are considered. However, in practice, designing a reflect beamforming vector needs to take some practical constraints into consideration, which inevitably limits the ability of RIS to mitigate interference. Take the maximum amplitude constraint for example, the solution in~\eqref{eq:a_cf} will be infeasible when some of REs are expected to have their amplitudes greater than $\alpha_{\max}$. For this case, the interference may be not completely mitigated, but the interference power level can still be minimized by solving the following constrained least squares problem 
\begin{align}
  \min_{\boldsymbol{a}}\quad & \left\|\boldsymbol{P}_\mathrm{I}^{\frac{1}{2}}\left(\boldsymbol{H}_\mathrm{b}^\mathrm{H} \boldsymbol{a} + \boldsymbol{h}_\mathrm{d}\right)\right\|^2 \label{prob:mini_interference}\\
  \mathrm{s.t.}\quad & \boldsymbol{a} \in \mathcal{V}, \nonumber
\end{align}
where $\boldsymbol{P}_\mathrm{I}=\mathrm{diag}([p_1,p_2,...,p_k]^\mathrm{T})\otimes \mathbf{I}_{K-1}$ represents the transmit power in a diagonal matrix, $\mathcal{V}$ represents the feasible set with these practical constraints on the RIS. Considering the maximum amplitude constraint first, we have
\begin{align}\label{eq:cst_a_amx}
  \mathcal{V} = \left\{ \boldsymbol{a} |\alpha_q \leq \alpha_{\max}, \forall q\right\}.
\end{align}
With the above constraints~\eqref{eq:cst_a_amx}, the interference power minimization problem can be recast into a quadratically constrained quadratic programming (QCQP) problem, which can be solved efficiently with the numerical interior-point method. 

\begin{figure}[t]
  \centering
  \includegraphics[width=.99\columnwidth]{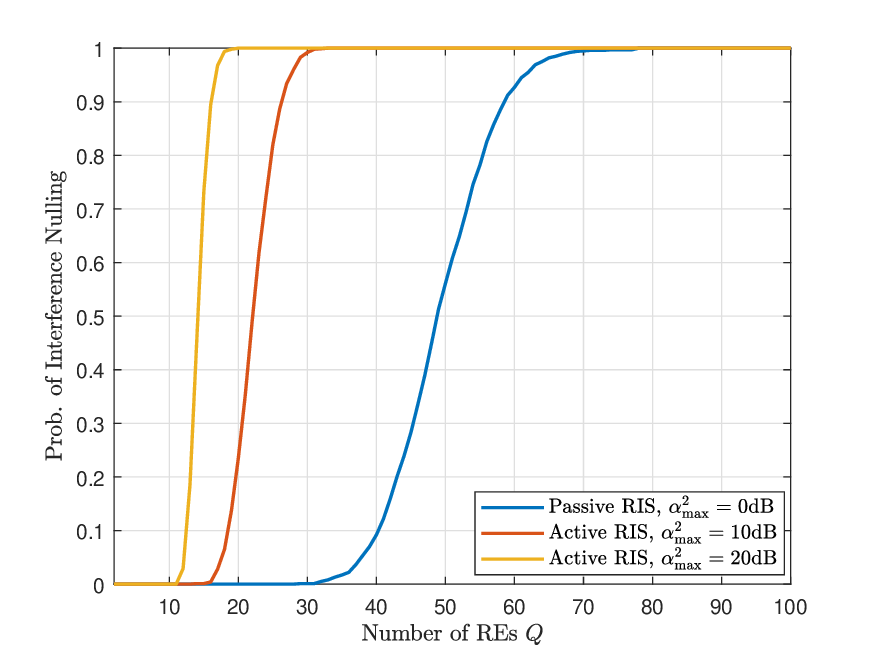}
  \caption{Minimize the interference power by solving problem~\eqref{prob:mini_interference} with the constraints~\eqref{eq:cst_a_amx}. Compared to the passive RIS, the active RIS use less REs to null the interference with high probability.}
  \label{fig:prob}
\end{figure}

To show how the maximum amplitude constraint affects the ability of RIS in mitigating interference, we consider an example of minimizing the interference power when there is no obstacle between the Tx and Rx pairs. Due to the double fading attenuation, the strength of the cross channel $h_{\mathrm{d},kj}$ is usually stronger than that of the cascaded channel $[\boldsymbol{h}_{\mathrm{b},kj}]_q$ of a single RE. Thus, we assume that $h_{\mathrm{b},kj}$, $\boldsymbol{h}_{\mathrm{t},j}$ and $\boldsymbol{h}_{\mathrm{r},k}$ are Rayleigh fading channels with $\mathbb{E}[|h_{\mathrm{d},kj}|^2]/\mathbb{E}[|[\boldsymbol{h}_{\mathrm{r},k}]_q|^2|[\boldsymbol{h}_{\mathrm{t},j}]_q|^2]=20\mathrm{dB}$, $p_k/\sigma_s^2 = 10\mathrm{dB}$, $\forall k$ and $K=4$.

Define the success probability of achieving interference nulling as follows
\begin{align}\label{eq:success_prob}
  \mathrm{Prob}\left\{\left\|\boldsymbol{P}_\mathrm{I}^{\frac{1}{2}}\left(\boldsymbol{H}_\mathrm{b}^\mathrm{H} \boldsymbol{a}^{*} + \boldsymbol{h}_\mathrm{d}\right)\right\|^2\leq .001\sigma_s^2\right\},
\end{align}
where $\boldsymbol{a}^{*}$ is the solution to \eqref{prob:mini_interference}. As shown in Fig.~\ref{fig:prob}, it is observed that when the strength of the direct channel is relatively strong, the passive RIS with $\alpha_{\max}=1$ needs a great quantity of REs to perfectly mitigate the interference, while the active RIS with a larger $\alpha_{\max}$ needs fewer REs to do so. Besides, as $\alpha_{\max}$ increases, the number of REs to null the interference approaches the number defined by the necessary condition $Q=K(K-1)$. This illustrates the significance of the applications of active RIS in interference channels.

\section{Power-aware Designs for Active RIS}\label{sec:powermodel}
As illustrated in the above section, the active RIS helps to suppress interference with a moderate number of REs in the presence of the strong cross  channels. In practice, the active RIS amplifies incident signals at the cost of a biasing power source for each RE. For a power-aware design, the power consumed by the active RIS is expected to be as little as possible to achieve specific goals. With the purpose of doing so, we first reinvestigate the power consumption model of the active RIS, and then propose power-aware designs for the active RIS.
\subsection{Power Consumption Model}
In~\cite{Long2021Active}, the power consumption model of the active RIS is given by
\begin{align}\label{eq:ori_power_model}
  P_\mathrm{aRIS}\! = \!Q(P_\mathrm{bias}\!+\!P_\mathrm{DC}) \!+\!  \xi (\|\boldsymbol{\Phi}\boldsymbol{H}_{\mathrm{t}}\boldsymbol{P}^{\frac{1}{2}}\|_{\mathrm{F}}^2 \!+ \!\sigma_r^2\|\boldsymbol{\Phi}\mathbf{I}_{Q}\|_{\mathrm{F}}^2),\!
\end{align}
which consists of the output power independent (OPI) component and the output power dependent (OPD) component. For the OPI component, $P_\mathrm{bias}$ is the biasing source power, and $P_\mathrm{DC}$ is the control circuit's operation power per RE. For the OPD component, $\xi$ is the amplification efficiency-related constant, $\boldsymbol{H}_{\mathrm{t}} = [\boldsymbol{h}_{\mathrm{t},1},\cdots,\boldsymbol{h}_{\mathrm{t},K}]$ is the forward channel matrix, and $\boldsymbol{P} = \mathrm{diag}([p_1,\cdots,p_K]^{\mathrm{T}})$ is the transmit power matrix. 

The above power consumption model has implicitly assumed that all the REs are identical and are tuned to an active mode to amplify incident signals. Intuitively, more active REs lead to better performances for active RIS-aided wireless systems, but it also results in a larger cost at the power consumption, because more biasing source power and circuit operation power are needed. As shown in Section~\ref{sec:Nulling}, the active RIS actually needs only parts of the REs to be activated for the purpose of mitigating the interference when each RE can provide high enough amplification gain. Hence, only the activated REs need to be adjusted to amplify incident signals, consuming additional biasing source power and circuit operation power. A more practical power consumption model can thus be given by
\begin{align}\label{eq:power_modified}
  p_\mathrm{aRIS} \!=\! \|\boldsymbol{\Phi}\|_0(P_\mathrm{bias} \!+\! P_\mathrm{DC}) \!+\! \xi (\|\boldsymbol{\Phi}\boldsymbol{H}_{\mathrm{t}}\boldsymbol{P}^{\frac{1}{2}}\|_\mathrm{F}^2 \!+\! \sigma_r^2\|\boldsymbol{\Phi}\mathbf{I}_{Q}\|_\mathrm{F}^2).
\end{align}
This model considers the sparsity of the reflect beamforming matrix whose non-zero entries correspond to the activated REs. By exploiting the power model in~\eqref{eq:power_modified}, we can maximize the utility of wireless systems while reducing the power consumption of the OPI components, leading to a greener design for active RIS.

For comparison, the power consumption model of the passive RIS is given by~\cite{Huang2019}
\begin{align}\label{eq:pRIS_power_model}
  p_\mathrm{pRIS} =Q P_\mathrm{DC}.
\end{align}
Since the passive RIS does not need to consume additional biasing and amplification power to support the signal amplification, their power consumption depends solely on the number of REs and the control circuit power per RE.

\subsection{Sum-Rate Maximization}
The interference power minimization proposed in~\eqref{prob:mini_interference} aims to find a zero-forcing solution to mitigate the interference which does not necessarily maximize the system utility. In the following, we consider a more meaningful problem that maximizes the sum rate of these user pairs in interference channels.

More specifically, with~\eqref{eq:power_modified}, a power budget constraint on the active RIS is introduced in the interference channels model, which can be rewritten as follows,
\begin{align}
  &\|\boldsymbol{a}\|_0(P_\mathrm{bias}+P_\mathrm{DC}) + \xi \boldsymbol{a}^\mathrm{H}\boldsymbol{E}_\mathrm{p}\boldsymbol{a} \leq P_\mathrm{RIS},\label{eq:power_budget}
\end{align}
where $\boldsymbol{E}_\mathrm{p}$ is a diagonal matrix with its diagonal elements being $\left[\boldsymbol{E}_{\mathrm{p}}\right]_{q,q} = \left[\boldsymbol{H}_\mathrm{t}\boldsymbol{P}\boldsymbol{H}_\mathrm{t}^\mathrm{H}\right]_{q,q}+\sigma_r^2$, $\forall q$ and $P_\mathrm{RIS}$ is the power budget for the RIS. It is worth mentioning that the active RIS can work with the power budget constraint~\eqref{eq:power_budget}, even when $P_\mathrm{RIS}$ is less than $Q(P_\mathrm{bias}+P_\mathrm{DC})$, because they are allowed to close some REs to meet the power budget with the model in~\eqref{eq:power_modified}. For comparison, the conventional model in~\eqref{eq:ori_power_model} cannot be applicable for this situation. 

Both the maximum amplitude constraint and the power-budget constraint are considered, and the sum-rate maximization problem is formulated as follows,
\begin{subequations}\label{prob:p1}
  \begin{align}
    \max_{\boldsymbol{a}}&~~~ \sum^K_{k=1} R_{\mathrm{a},k}(\boldsymbol{a}) \label{eq:p1obj}\\
    \mathrm{s.t.}&~~~ \|\boldsymbol{a}\|_0(P_\mathrm{bias}+P_\mathrm{DC}) + \xi \boldsymbol{a}^\mathrm{H}\boldsymbol{E}_\mathrm{p}\boldsymbol{a} \leq P_\mathrm{RIS},\label{eq:p1c1}\\
    &~~~\alpha_q \leq \alpha_{\max}, \forall q. \label{eq:p1c2}
  \end{align}
\end{subequations}

Problem~\eqref{prob:p1} is nonconvex due to the nonconvex objective function and nonconvex $\ell_0$-norm constraint in~\eqref{eq:p1c1}. More specifically, the $\ell_0$-norm constraint forces the active RIS to select parts of REs to be activated, which makes the problem become a NP-hard combinatorial problem. In general, it is thus quite challenging to find a global optimal solution to~\eqref{prob:p1} within an acceptable complexity.

\subsection{Active RIS Power Minimization}
In addition to the sum-rate maximization in interference channels, it is also of great interest to investigate the minimum power budget that the active RIS needs to guarantee a predetermined achievable rate for each user pair. To achieve these rate requirements, the active RIS directly reconfigures the channel conditions via optimizing the reflect beamforming vector, instead of the users allocating their transmit power in a collaborative manner. More specifically, the minimum rate constraint for each user pair and the maximum amplitude constraint are considered, and the active RIS power minimization problem is formulated as follows,
\begin{subequations}\label{prob:p2}
  \begin{align}
    \min_{\boldsymbol{a}}&~~~ \|\boldsymbol{a}\|_0(P_\mathrm{bias}+P_\mathrm{DC}) + \xi \boldsymbol{a}^\mathrm{H}\boldsymbol{E}_\mathrm{p}\boldsymbol{a} \label{eq:p2obj}\\
    \mathrm{s.t.}&~~~R_{\mathrm{a},k}(\boldsymbol{a}) \geq {R}_{k}, \forall k \label{eq:p2c1} \\
    &~~~ \alpha_q \leq \alpha_{\max}, \forall q. \label{eq:p2c2}
  \end{align}
\end{subequations}
Solving Problem~\eqref{prob:p2} will provide a straightforward power-aware design for the active RIS, in which the power budget is optimized directly by reducing the number of the activated REs, provided that these rate requirements are fulfilled. However, Problem~\eqref{prob:p2} is difficult to solve due to the nonconvex $\ell_0$-norm objective function and the nonconvex minimum rate constraints in~\eqref{eq:p2c2}. 

Notice that Problem~\eqref{prob:p2} may be infeasible, because such stringent achievable rate requirements cannot be satisfied with the current channel realizations. However, as shown later by numerical results, the feasibility of~\eqref{prob:p2} can be improved with the assistance of the active RIS. 

\section{The $\ell_0$-norm Relaxation}\label{sec:NormRelaxation}
Motivated by the compressive sensing literature,  we propose sparse reflect beamforming design algorithms to address both the sum-rate maximization problem~\eqref{prob:p1} and the active RIS power minimization problem~\eqref{prob:p2}. The algorithms primarily focus on relaxing the $\ell_0$ norm in~\eqref{eq:power_budget} to a weighted $\ell_1$ norm. This is achieved by iteratively updating the weights and solving a sequence of problems. Initially, the $\ell_0$-norm is relaxed to facilitate the solution of both problem types.

As the $\ell_0$ norm is defined by the number of non-zero entries of a vector, it is observed that 
\begin{align}\label{eq:l0_norm}
  \|\boldsymbol{a}\|_0 = \|[|\alpha_1|^2,\cdots,|\alpha_Q|^2]^\mathrm{T}\|_0.
\end{align}
Inspired by~\cite{candes2008enhancing}, the right-hand side of~\eqref{eq:l0_norm} can be relaxed with a weighted $\ell_1$ norm as follows
\begin{align}
 \|[|\alpha_1|^2,\cdots,|\alpha_Q|^2]^\mathrm{T}\|_0 =  \|[\beta_1|\alpha_1|^2,\cdots,\beta_Q|\alpha_Q|^2]^\mathrm{T}\|_1, \nonumber
\end{align}
where $\beta_q,~q = 1,\cdots,Q $ is a positive weight associated with the amplification gain $|\alpha_q|^2$ of the $q$-th RE. By applying the $\ell_0$-norm relaxation into the proposed power consumption model~\eqref{eq:power_modified}, it can be rewritten by 
\begin{align}
  P_\mathrm{aRIS}&=(P_\mathrm{bias}\!+\!P_\mathrm{DC})\sum_{q=1}^{Q}\beta_q|\alpha_q|^2 \!+\!\xi \boldsymbol{a}^\mathrm{H}\boldsymbol{E}_\mathrm{p}\boldsymbol{a}\! \nonumber\\
  & = \!\boldsymbol{a}^\mathrm{H} \boldsymbol{E}_\mathrm{a} \boldsymbol{a}, \label{eq:power_budget_cst}
\end{align}
where $\boldsymbol{E}_\mathrm{a}$ is a diagonal matrix with its diagonal elements being
\begin{align}
  \left[\boldsymbol{E}_\mathrm{a}\right]_{q,q} = (P_\mathrm{bias}+P_\mathrm{DC})\beta_q + \xi \left[\boldsymbol{E}_\mathrm{p}\right]_{q,q}, \forall q.
\end{align}
Provided with the $\ell_0$-norm relaxation and the appropriate weights, the power consumption model for the active RIS can be represented by a convex quadratic function, which facilitates solving the considered problems. However, it is still a challenging task to find the appropriate weights. Based on the heuristic updating rule in~\cite{candes2008enhancing,Dai2014Sparse}, the weights $\beta_q$ can be iteratively updated according to
\begin{align}\label{eq:weights}
  \beta_q = \frac{1}{|\alpha_q|^2+\tau},\quad \forall q,
\end{align}
where $|\alpha_q|^2$ is obtained in previous iteration and $\tau$ is a small constant. For ease of explanation, a vector $\boldsymbol{\beta}$ is used to denote the vector that collects all the weights $\beta_q, \forall q$. In~\eqref{eq:weights}, the weights are designed to be inversely proportional to the true amplification gain $|\alpha_q|^2$, which makes those REs with a lower amplification gain would have higher weights. This scheme forces the active RIS to eventually close those REs to save power. In addition, the small constant $\tau$ is introduced to provide stability and to ensure that the denominator in~\eqref{eq:weights} is non-zero for a zero-valued $|\alpha_q|$. It is shown numerically later that this updating rule promotes the sparsity in the design for the active RIS.

\section{Sparse Reflect Beamforming Design Algorithm}\label{sec:Algs}
In this section, we propose two distinct algorithms for sparse reflect beamforming design, each designed to address specific optimization challenges: one for maximizing the sum rate and another for minimizing the active RIS power consumption.

\subsection{Proposed Algorithm for Sum-Rate Maximization}
Although the above relaxation provides a way to tackle the nonconvex constraint by replacing~\eqref{eq:power_budget} with~\eqref{eq:power_budget_cst}, the sum-rate maximization problem cannot be directly solved in the current form. This is because, in the objective function of~\eqref{prob:p1}, the SINR ratio terms that reside in the logarithm functions are in fractional form with respective to $\boldsymbol{a}$. Such a sum-of-logarithms-of-ratio problem can be tackled by fractional programming in~\cite{Shen2018Fractional}. More specifically, by introducing a set of auxiliary variables $\omega_k \in \mathbb{C}, \forall k$, Problem~\eqref{prob:p1} can be rewritten as
\begin{subequations}\label{prob:fp}
  \begin{align}
    \max_{\boldsymbol{a},\boldsymbol{\omega}}~ &\sum_{k=1}^{K}\log\left(1+\sqrt{p_k}\mathrm{Re}\left(\omega_k^{*}(h_{\mathrm{d},kk} + \boldsymbol{h}^{\mathrm{H}}_{\mathrm{b},kk}\boldsymbol{a})\right)\right. \nonumber\\
    & \left.-\omega_k^{*}(\boldsymbol{a}^\mathrm{H}\left(\boldsymbol{R}_{\mathrm{r},k}+\boldsymbol{R}_{\mathrm{b},k}\right)\boldsymbol{a}+2\mathrm{Re}(\boldsymbol{g}_{k}^{\mathrm{H}}\boldsymbol{a})+C_k)\omega_k \right), \\
    \mathrm{s.t.}~& \alpha_q \leq \alpha_{\max}, \forall q,\\
    & \boldsymbol{a}^\mathrm{H} \boldsymbol{E}_\mathrm{a} \boldsymbol{a} \leq P_\mathrm{RIS}
  \end{align}
\end{subequations}
where $\boldsymbol{\omega}$ is a vector that collects all the auxiliary variables $\omega_k,~\forall k$, $\boldsymbol{R}_{\mathrm{r},k}= \sigma_r^2\mathrm{diag}(|\boldsymbol{h}_{\mathrm{r},k}|^2)$, $\boldsymbol{R}_{\mathrm{b},k}=\sum_{j\neq k}^{K}p_j\boldsymbol{h}_{\mathrm{b},kj}\boldsymbol{h}_{\mathrm{b},kj}^{\mathrm{H}}$, and $C_k = \sum_{j\neq k}^{K}p_j|h_{\mathrm{d},kj}|^2+\sigma_s^2$. 

Notice that given $\boldsymbol{a}$, Problem~\eqref{prob:fp} is a convex problem with respective to $\boldsymbol{\omega}$, and the objective function is maximized when 
\begin{align}\label{eq:omega_opt}
  \omega_k = \frac{\sqrt{p_k}(h_{\mathrm{d},kk}+\boldsymbol{h}^{\mathrm{H}}_{\mathrm{b},kk}\boldsymbol{a})}{\boldsymbol{a}^\mathrm{H}\left(\boldsymbol{R}_{\mathrm{r},k}+\boldsymbol{R}_{\mathrm{b},k}\right)\boldsymbol{a}+2\mathrm{Re}(\boldsymbol{g}_{k}^{\mathrm{H}}\boldsymbol{a})+C_k},\quad \forall k.
\end{align}
Besides, given $\boldsymbol{\omega}$, Problem~\eqref{prob:fp} is also a convex problem, and thus the optimal $\boldsymbol{a}$ can be obtained by solving the convex problem with an efficient convex optimization tool like CVX~\cite{grant2008cvx}. By iteratively updating $\boldsymbol{\omega}$ and $\boldsymbol{a}$, a high-quality stationary point to \eqref{prob:fp} can be obtained.

\begin{algorithm}[t!]
  \caption{Fractional programming}\label{alg:fp}
  \begin{algorithmic}[1]
  \STATE Initialize $\boldsymbol{a}$ feasible values. \\
  \REPEAT
  \STATE Update $\boldsymbol{\omega}$ by~\eqref{eq:omega_opt}.
  \STATE Update $\boldsymbol{a}$ by solving \eqref{prob:fp} with fixed $\boldsymbol{\omega}$.
  \UNTIL{Convergence} 
  \end{algorithmic}
\end{algorithm}

So far, we have successfully solved~\eqref{prob:fp} with given $\boldsymbol{\beta}$ by solving a fractional programming problem in an iterative manner. After obtaining the solution $\boldsymbol{a}$, we update the weights $\boldsymbol{\beta}$ based on~\eqref{eq:weights} for the next iteration until convergence. Finally, we summarize the proposed algorithm in Algorithm~\ref{alg:SRB}.

\begin{algorithm}[t!]
  \caption{Two-loop sparse reflect beamforming design for sum-rate maximization}\label{alg:SRB}
  \begin{algorithmic}[1]
  \STATE Initialize $\boldsymbol{a}$ feasible values by solving Problem~\eqref{prob:mini_interference}. \\
  \REPEAT
  \STATE Update the weights $\boldsymbol{\beta}$ based on~\eqref{eq:weights}.
  \STATE Update the reflect beamforming $\boldsymbol{\alpha}$ with fractional programming.
  \UNTIL{Convergence} 
  \end{algorithmic}
\end{algorithm}
Solving the interference power minimization problem~\eqref{prob:mini_interference} with only the maximum amplitude constraint produces a good initial point that helps the fractional programming method to avoid undesirable local maxima. As the iteration in Algorithm~\ref{alg:SRB} goes on, the active RIS eventually finds the sparse reflect beamforming vector of which parts of REs have their amplitude close to zero. 

For the REs whose amplitude is close to zero, the active RIS needs not activate them with additional power but leave them to a passive state which can be achieved with passive RIS designs. However, in this paper, we pay more attention to the performance improvement brought by the activated REs with biasing sources, and thus we set the REs with the amplitude less than unity to be closed. 

\begin{algorithm}[t!]
  \caption{One-loop sparse reflect beamforming design for sum-rate maximization}\label{alg:SRB_low}
  \begin{algorithmic}[1]
  \STATE Initialize $\boldsymbol{a}$ feasible values. \\
  \REPEAT
  \STATE Update the weights $\boldsymbol{\beta}$ based on~\eqref{eq:weights};
  \STATE Update $\boldsymbol{\omega}$ by~\eqref{eq:omega_opt}.
  \STATE Update $\boldsymbol{a}$ by solving \eqref{prob:fp} with fixed $\boldsymbol{\beta}$ and $\boldsymbol{\omega}$.
  \UNTIL{Convergence} 
  \end{algorithmic}
\end{algorithm}

Algorithm~\ref{alg:SRB} is proposed to obtain the sparse reflect beamforming vector with two loops, namely an inner loop to solve the $\ell_1$-norm sum-rate maximization problem~\eqref{prob:fp}  with given $\boldsymbol{\beta}$, and an outer loop to update the weights. Although the inner loop is a standard fractional programming algorithm whose convergence can be guaranteed by obtaining a stationary point~\cite{Shen2018Fractional}, this two-loop algorithm can have a high computational complexity. To reduce the complexity, a one-loop algorithm is proposed that directly update $\boldsymbol{\beta}$ inside the inner loop of the fractional programming algorithm~\eqref{alg:fp}, instead of updating it after the finalization of the inner loop, as summarized in Algorithm~\ref{alg:SRB_low}. The convergence of Algorithm~\ref{alg:SRB_low} is later verified by the numerical simulation.

\emph{Complexity analysis:} The complexity of the algorithms for the sum-rate maximization mainly depends on solving the FP problem~\eqref{prob:fp} with the fixed $\boldsymbol{\omega}$ in each iteration. The objective function in~\eqref{prob:fp} is a sum of logarithms with the concave quadratic function composited in the logarithm function, and it can be recast into the product of these concave quadratic functions. Since the concave quadratic functions are all second-order cone (SOC)-representable, the objective function in~\eqref{prob:fp} is readily expressible as a system of SOC constraints~\cite{lobo1998applications}. Thus, solving Problem~\eqref{prob:fp} is in fact equivalent to solving an SOC programming (SOCP) problem with $Q$ variables and $Q+1$ constraints. With the primal-dual interior point method, the number of iterations needed to decrease the dual gap to a constant fraction of itself is upper bounded by $\mathcal{O}\left(\sqrt{Q+1}\right)$, and for each iteration the worst-case complexity is $\mathcal{O}\left(Q^2\right)$. Therefore, for each iteration, the complexity of solving the FP problem given $\boldsymbol{\omega}$ is $\mathcal{O}(Q^{3.5})$. 

For the two-loop algorithm, the total complexity of solving the sum-rate maximization problem is $\mathcal{O}(I_\mathrm{two}I_\mathrm{fp}Q^{3.5})$, where $I_\mathrm{two}$ is the number of iterations for the outer loop of the weight updates to converge, and $I_\mathrm{fp}$ is the number of iterations for the inner loop of FP to converge. Likewise, for the one-loop algorithm, the total complexity of solving the sum-rate maximization problem is $\mathcal{O}(I_\mathrm{one}Q^{3.5})$, where $I_\mathrm{one}$ is the number of iterations for the one-loop algorithm to converge. As shown in the later simulation results, these two kinds of algorithms can obtain a stationary solution with $I_\mathrm{two}$ and $I_\mathrm{one}$ almost the same. Additionally, due to the lack of an inner loop, the one-loop algorithm has less computational complexity.

\subsection{Proposed Algorithm for Active RIS Power Minimization}
The $\ell_0$-norm relaxation provides the way to convert the nonconvex objective function of~\eqref{prob:p2} into the tractable quadratic function, whereas the rate requirement constraints~\eqref{eq:p2c1} are still nonconvex. To address this problem, we first recast the problem into a semidefinite programming (SDP) problem after the semidefinite relaxation (SDR), and then recover the rank-one solution with difference of convex (DC) programming. 

The rate requirement constraints are equivalent to the following SNR constraints
\begin{align}
  \frac{p_k \left|h_{\mathrm{d},kk} + \boldsymbol{h}^{\mathrm{H}}_{\mathrm{b},kk}\boldsymbol{a}\right|^2}{\boldsymbol{a}^\mathrm{H}\left(\boldsymbol{R}_{\mathrm{r},k}+\boldsymbol{R}_{\mathrm{b},k}\right)\boldsymbol{a}+2\mathrm{Re}(\boldsymbol{g}_{k}^{\mathrm{H}}\boldsymbol{a})+C_k} \geq \gamma_k,~\forall k, \label{eq:snr_cst}
\end{align}
with $\gamma_k = 2^{R_k}-1$. Let $\bar{\boldsymbol{a}} =[\boldsymbol{a};a_{Q+1}]\in \mathbb{C}^{(Q+1)\times 1}$, where $a_{Q+1}$ is an auxiliary variable with $a_{Q+1} = 1$. The SNR constraints are rewritten as
\begin{align}
  \frac{\mathrm{Tr}\left(\boldsymbol{R}_{kk}\bar{\boldsymbol{a}}\bar{\boldsymbol{a}}^\mathrm{H}\right)}{\mathrm{Tr}\left(\boldsymbol{R}_{\mathrm{rb},k}\bar{\boldsymbol{a}}\bar{\boldsymbol{a}}^\mathrm{H}\right)}\geq \gamma_k,~\forall k,
\end{align}
where
\begin{align*}
  \boldsymbol{R}_{kk}&= p_k
    \left[
    \begin{array}{cc}
   \boldsymbol{h}_{\mathrm{b},kk}\boldsymbol{h}_{\mathrm{b},kk}^\mathrm{H} & \boldsymbol{h}_{\mathrm{b},kk}h_{\mathrm{d},kk}^{*} \\
   h_{\mathrm{d},kk}\boldsymbol{h}_{\mathrm{b},kk}^\mathrm{H} & |h_{\mathrm{d},kk}|^2 \\
  \end{array}
  \right],\\
  \boldsymbol{R}_{\mathrm{rb},k}&=
  \left[
    \begin{array}{cc}
    \boldsymbol{R}_{\mathrm{r},k}+\boldsymbol{R}_{\mathrm{b},k} &  \boldsymbol{g}_k \\
    \boldsymbol{g}_k^\mathrm{H} & C_k \\
  \end{array}
  \right].
\end{align*}
Then, a positve semidefinite matrix variable $\boldsymbol{A}=\bar{\boldsymbol{a}}\bar{\boldsymbol{a}}^\mathrm{H}$ is introduced, and the original problem~\eqref{prob:p2} can be recast into the problem as follows,
\begin{subequations}\label{prob:p2-sdp}
  \begin{align}
    \min_{\boldsymbol{A}} \quad & \mathrm{Tr}\left(\overline{\boldsymbol{E}}_\mathrm{a}\boldsymbol{A}\right) \label{eq:p2-sdp_obj}\\
    \mathrm{s.t.}\quad& {\mathrm{Tr}\left(\boldsymbol{R}_{kk}\boldsymbol{A}\right)}-\gamma_k{\mathrm{Tr}\left(\boldsymbol{R}_{\mathrm{rb},k}\boldsymbol{A}\right)}\geq 0,~\forall k \label{eq:p2_sdp_c1} \\
    &\mathrm{Tr}\left(\overline{\boldsymbol{E}}_{q}\boldsymbol{A}\right) \leq \alpha_{\max}^2, q = 1,\cdots, Q, \\
    &\mathrm{Tr}\left(\overline{\boldsymbol{E}}_{Q+1}\boldsymbol{A}\right) = 1, \label{eq:p2-sdp_unit}\\
    &\boldsymbol{A}\succcurlyeq 0 \label{eq:p2_sdp_c4}\\
    &\mathrm{Rank}(\boldsymbol{A}) = 1. \label{eq:p2-sdp_rank}
  \end{align}
\end{subequations}
where $\overline{\boldsymbol{E}}_\mathrm{a}$ is a block diagonal matrix with $\overline{\boldsymbol{E}}_\mathrm{a} = \mathrm{blkdiag}\left(\boldsymbol{E}_\mathrm{a},0\right)$, and $\overline{\boldsymbol{E}}_{q}$ is a square matrix with the $q$-th diagonal element being $1$ and the other elements being $0$ for $q = 1,\cdots,Q+1$. Without the rank-one constraint~\eqref{eq:p2-sdp_rank}, Problem~\eqref{prob:p2-sdp} is an SDP problem which can be efficiently solved with the convex optimization tool.

Consider the solution to the SDP problem is denoted by $\boldsymbol{A}^{*}$. If $\mathrm{Rank}(\boldsymbol{A}^{*})=1$, the reflect beamforming vector can be obtained with $\boldsymbol{a}^{*} = \bar{\boldsymbol{a}}^{*}_{[1:Q]}$ where $\bar{\boldsymbol{a}}^{*}$ can be obtained with Cholesky decomposition of $\boldsymbol{A}^{*}=\bar{\boldsymbol{a}}^{*}(\bar{\boldsymbol{a}}^{*})^\mathrm{H}$. Otherwise, the techniques to generate the rank-one solution with $\boldsymbol{A}^{*}$ are needed. Conventionally, the rank-one solution can be recovered with the Gaussian randomization method, which is widely adopted by the designs with the passive RIS~\cite{Wu2019}. However, such a Gaussian randomization method cannot promise to find a rank-one suboptimal solution that simultaneously fulfills the requirements in Problem~\eqref{prob:p2-sdp}, especially for~\eqref{eq:p2-sdp_unit}.

Instead of the Gaussian randomization method, the DC programming method is adopted in this paper to recover the rank-one solution. The main idea of the proposed DC method is that the rank constraint~\eqref{eq:p2-sdp_rank} can be replaced with a difference-of-convex constraint as follows,
\begin{align}\label{eq:}
  \mathrm{Tr}\left(\boldsymbol{A}\right)-\left\|\boldsymbol{A}\right\|_2 = 0,
\end{align}
of which the left-hand side is a continuous function. The replacement is valid when $\boldsymbol{A}$ is a positive semidefinite matrix~\cite{Yang2020Federated}. The constraint can then be added as a penalty term at the objective function~\eqref{eq:p2-sdp_obj}. In this way, Problem~\eqref{prob:p2-sdp} is rewritten as
\begin{subequations}\label{eq:p2_penalty}
\begin{align}
  \min_{\boldsymbol{A}}\quad& \mathrm{Tr}\left(\overline{\boldsymbol{E}}_\mathrm{a}\boldsymbol{A}\right) + \rho \left(\mathrm{Tr}\left(\boldsymbol{A}\right)-\left\|\boldsymbol{A}\right\|_2\right) \label{eq:penalty_term}\\
  \mathrm{s.t.}\quad & \boldsymbol{A}\in \mathcal{F}_{p},
\end{align}
\end{subequations}
where $\rho\geq 0$ is the penalty factor which enforces the penalty component in~\eqref{eq:penalty_term} to be zero, and $\mathcal{F}_{p}$ is the convex feasible set characterized by the constraints~\eqref{eq:p2_sdp_c1}-\eqref{eq:p2_sdp_c4}. Let
\begin{align}
  g(\boldsymbol{A}) &= \mathrm{Tr}\left(\overline{\boldsymbol{E}}_\mathrm{a}\boldsymbol{A}\right) + \rho\mathrm{Tr}\left(\boldsymbol{A}\right), \\
  h(\boldsymbol{A}) &= \rho \left\|\boldsymbol{A}\right\|_2,
\end{align}
the original problem is thus represented by
\begin{subequations}\label{prob:p2-dc}
  \begin{align}
    \min_{\boldsymbol{A}}&\quad f(\boldsymbol{A}) = g(\boldsymbol{A}) - h(\boldsymbol{A}), \\
    \mathrm{s.t.}&\quad \boldsymbol{A}\in \mathcal{F}_{p},
  \end{align}
\end{subequations}
whose objective function is presented in the form of the DC function. Although the DC function is still nonconvex, a high-quality local optimal solution can be obtained by successively solving the convex relaxation of primal and dual problems of~\eqref{prob:p2-dc}, which is based on the basic principles of the DC algorithm (DCA) in~\cite{le2018dc}. More specifically, for each iteration $t$, the second DC component $h(\boldsymbol{A})$ is approximated by its affine minorization shown below
\begin{align}\label{eq:h_minorant}
  h_{t}(\boldsymbol{A}) = h(\boldsymbol{A}^{t}) + \left\langle \boldsymbol{A}-\boldsymbol{A}^{t},\boldsymbol{Z}^{t}\right\rangle,
\end{align}
where $\left\langle \boldsymbol{X},\boldsymbol{Y}\right\rangle = \mathrm{real}\left(\mathrm{Tr}\left(\boldsymbol{X}^\mathrm{H}\boldsymbol{Y}\right)\right)$ is the inner product of two matrices, $\boldsymbol{A}^{t}$ and $\boldsymbol{Z}^{t}$ are the solutions to convex relaxation of the primal and the dual problem in the $t$-th iteration, respectively, and $\boldsymbol{Z}^{t} \in \partial h(\boldsymbol{A}^{t})$. With the affine minorization, the local optimal $\boldsymbol{A}^{*}$ is obtained with the update over the sequence $\left\{\boldsymbol{A}^{t}\right\}$. The update rule follows 
\begin{align}
  \boldsymbol{Z}^{t} & = \arg\min_{\boldsymbol{Z}\in\mathcal{F}_{d}}\left\{h^{*}(\boldsymbol{Z})-g^{*}_{t-1}(\boldsymbol{Z})\right\}, \label{eq:update_dual}\\
  \boldsymbol{A}^{t+1} & = \arg\min_{\boldsymbol{A}\in\mathcal{F}_{p}}\left\{g(\boldsymbol{A})-h_{t}(\boldsymbol{A})\right\}, \label{eq:A_update}
\end{align}
where $h^{*}$ and $g^{*}$ are the conjugate function of $h$ and $g$, respectively, and $\mathcal{F}_{d}$ is the dual set of $\mathcal{F}_{p}$. The conjugate function is defined by 
\begin{align}
  g^{*}(\boldsymbol{Z}) = \sup \left\{\left\langle\boldsymbol{A},\boldsymbol{Z}\right\rangle - g(\boldsymbol{A})\right\}.
\end{align}
$g^{*}_{t-1}(\boldsymbol{Z})$ is the majorant of $g^{*}$ at $\boldsymbol{Z}^{t-1}$, which is given by
\begin{align}
  g^{*}_{t-1}(\boldsymbol{Z}) = g^{*}(\boldsymbol{Z}^{{t-1}}) - \left\langle\boldsymbol{Z}-\boldsymbol{Z}^{{t-1}},\boldsymbol{A}^{t} \right\rangle,
\end{align}
with $\boldsymbol{A}^{t} \in \partial g^{*}(\boldsymbol{Z}^{t-1})$. In addition to the update on the sequence $\left\{\boldsymbol{A}^{t}\right\}$, it is necessary to update the sequence $\boldsymbol{Z}^{t}$ to implement DCA. Instead of solving the dual problem~\eqref{eq:update_dual}, one direct method to update $\boldsymbol{Z}^{t}$ is to calculate the subgradient of $h$ at $\boldsymbol{A}^{t-1}$, since $\boldsymbol{Z}^{t}\in\partial h(\boldsymbol{A}^{t})$ accodring to the Fenchel biconjugation theorem~\cite{rockafellar2015convex}. As $h(\boldsymbol{A})=\left\|\boldsymbol{A}\right\|_2$ is the spectral norm of $\boldsymbol{A}$, one of its subgradient can be calculated with~\cite{Yang2020Federated}
\begin{align}\label{eq:sub_grad}
  \boldsymbol{u}_{1}\boldsymbol{u}_{1}^\mathrm{H} \in \partial h(\boldsymbol{A}),
\end{align}
where $\boldsymbol{u}_{1}\in \mathbb{C}^{(Q+1)\times 1}$ is the eigenvector of the largest eigenvalue $\sigma_1(\boldsymbol{A})$. Therefore, $\boldsymbol{Z}^{t}$ in~\eqref{eq:h_minorant} can be updated by the following equation
\begin{align}\label{eq:Z_update}
  \boldsymbol{Z}^{t} = \boldsymbol{u}^{t}_{1}\left(\boldsymbol{u}^{t}_{1}\right)^\mathrm{H}.
\end{align}

With the affine relaxation in~\eqref{eq:h_minorant}, Problem~\eqref{eq:A_update} is a convex problem which can be efficiently solved with the convex optimization tool. The proposed algorithm that exploits the DCA to recover the rank-one solution is concluded in Algorithm~\ref{alg:p2_sdp}. 

By now, given the weights of the $\ell_1$-norm relaxation $\boldsymbol{\beta}$, the reflect beamforming vector $\boldsymbol{a}$ to minimize the active RIS power consumption is obtained with the proposed iterative primal-dual subgradient DCA. In order to promote the sparsity of the solution, given $\boldsymbol{a}$, $\boldsymbol{\beta}$ needs to be iteratively updated with~\eqref{eq:weights} for the next iteration until convergence. The complete algorithm for the power minimization problem is summarized in Algorithm~\ref{alg:SRB_apm}. Notice that the initialization can still be achieved by solving the interference minimization problem~\eqref{prob:mini_interference}, since the initialization can affect the objective function but cannot affect the feasible region of Problem~\eqref{prob:p2}. The feasibility of the considered power minimization problem can be verified by solving the  SDP problem~\eqref{prob:p2-sdp} in the initialization of Algorithm~\ref{alg:p2_sdp}. If the initialization fails, claim Problem~\eqref{prob:p2} is infeasible under the current rate requirements. The local convergence of the considered DCA can be proved with the strongly convex functions representation~\cite{Yang2020Federated}.

\begin{algorithm}[t!]
  \caption{Algorithm to recover the rank-one solution to Problem~\eqref{prob:p2-sdp}.}\label{alg:p2_sdp}
  \begin{algorithmic}[1]
  \STATE Initialize $\boldsymbol{A}^{*}$ by solving the SDP Problem~\eqref{prob:p2-sdp} without the rank-one constraint. \\
  \IF{$\mathrm{Rank}(\boldsymbol{A}^{*}) \neq 1$.} 
  \STATE {$\boldsymbol{A}^{t} = \boldsymbol{A}^{*}$, and set $t = 1$.} \\
  \REPEAT 
  \STATE {Update $\boldsymbol{Z}^{t}$ with~\eqref{eq:Z_update}.}
  \STATE {Update $\boldsymbol{A}^{t+1}$ by solving Problem~\eqref{eq:A_update}.}
  \UNTIL{The penalty component of Problem~\eqref{eq:p2_penalty} is below a small threshold.}  
  \STATE $\boldsymbol{A}^{*} = \boldsymbol{A}^{t+1}$.
  \ENDIF
  \STATE $\boldsymbol{a}^{*} = \bar{\boldsymbol{a}}^{*}_{[1:Q]}/\bar{\boldsymbol{a}}^{*}_{[Q+1]}$ where $\bar{\boldsymbol{a}}^{*}$ can be obtained with Cholesky decomposition of $\boldsymbol{A}^{*}=\bar{\boldsymbol{a}}^{*}(\bar{\boldsymbol{a}}^{*})^\mathrm{H}$.
  \end{algorithmic}
\end{algorithm}

\begin{algorithm}[t!]
  \caption{Sparse reflect beamforming design for active RIS power minimization}\label{alg:SRB_apm}
  \begin{algorithmic}[1]
  \STATE Initialize $\boldsymbol{a}$ feasible values by solving Problem~\eqref{prob:mini_interference}. \\
  \REPEAT
  \STATE Update the weights $\boldsymbol{\beta}$ based on~\eqref{eq:weights}.
  \STATE Update the reflect beamforming $\boldsymbol{\alpha}$ with the proposed DCA in Algorithm~\ref{alg:p2_sdp}.
  \UNTIL{Convergence} 
  \end{algorithmic}
\end{algorithm}

\emph{Complexity analysis:} The complexity of the complete algorithm for the active RIS power minimization mainly lies in solving the SDP problem~\eqref{eq:p2_penalty} in each iteration. The dimensions of the input variables and the constraints on the SDP problem affect the computational complexity. Since the considered SDP problem has a $(Q+1)\times(Q+1)$ PSD matrix variable and $(K+Q+1)$ PSD constraints, it usually takes $\mathcal{O}\left(\sqrt{Q+1}\right)$ iterations to decrease the dual gap of the desired accuracy, and the worst case complexity $\mathcal{O}\left((K+Q+1)^4\right)$ for each iteration in the interior point method~\cite{Luo2010}. Therefore, the complexity of solving the SDP problem is $\mathcal{O}\left(\sqrt{Q+1}(K+Q+1)^4\right)$ during an iteration. Consider the number of iterations to recover the rank-one solution is $I_\mathrm{p}$, the complexity of Algorithm~\ref{alg:p2_sdp} is $\mathcal{O}\left((I_\mathrm{p}+1)\sqrt{Q+1}(K+Q+1)^4\right)$, where the SDP problem has to be solved for one time at least. Moreover, to obtain the sparse reflect beamforming vector, the reweighted factor $\boldsymbol{\beta}$ should be updated in $I_{\mathrm{re}}$ iterations. The computational complexity of the complete algorithm in Algorithm~\ref{alg:SRB_apm} is thus $\mathcal{O}\left(I_{\mathrm{re}}(I_\mathrm{p}+1)\sqrt{Q+1}(K+Q+1)^4\right)$.

\section{Comparisons with Conventional RIS Designs}\label{sec:comparison}
In this study, we have investigated the sum-rate maximization and power minimization in active RIS-aided interference channels. For the sake of energy saving, we have proposed the power-aware sparse reflect beamforming designs on active RIS, which allow it to flexibly use its power budget by closing parts of the inefficient REs that suffer from the poor channel conditions on signal propagations. An important aspect of our research is to assess the energy-saving potential of these power-aware designs in comparison to traditional RIS designs.

\subsubsection{Fully Active RIS}
The fully active RIS design, where the active RIS activates all the REs to amplify incident signals, offers the best performance of maximizing the sum rate when no power budget is taken into account. However, this design generally leads to higher power consumption, due to the fact that activating all the REs needs to consume plenty of additional biasing power. In scenarios where sufficient power is available at the RIS, fully utilizing all REs is advantageous for significant performance enhancement through signal amplification. The performance of this design in interference channels can be effectively analyzed using our proposed algorithms, by adapting the power consumption model as described in~\eqref{eq:ori_power_model}.

\subsubsection{Fixed Active RIS}
The fixed active RIS design involves activating a predetermined number of REs under the constraint of a limited power budget. Specifically, the power budget limits that only $Q_{F} = \min(\lfloor\frac{P_\mathrm{RIS}}{P_\mathrm{bias}+P_\mathrm{DC}}\rfloor,Q)$ REs can be activated. This necessitates selecting $Q_{F}$ REs from the total available $Q$ REs. Addressing this selection as a combinatorial problem, represented as $Q \choose Q_{F}$, becomes computationally challenging as $Q$ increases. A simpler, though less optimized, approach is to activate a fixed set of $Q_{F}$ REs, such as the first or last $Q_{F}$ REs in the surface. While this strategy may lead to some performance degradation, it offers a practical solution for operating active RIS under limited power resources. The performance of this design in interference channels can be analyzed using our proposed algorithms, which involve adapting the power consumption model as per~\eqref{eq:ori_power_model} for the specified $Q_{F}$ activated REs.

\subsubsection{Passive RIS}
The passive RIS design leverages a significant number of REs without signal amplification capabilities. Within a given power budget, passive RIS can support a considerably larger number of REs compared to its active counterpart. This is because REs in a passive RIS consume only control circuit power and does not need for additional biasing sources, thereby reducing the power cost. Additionally, the absence of active RF components in passive RIS means that no additional noise is introduced during signal reflection. Notably, the signal power reflected through passive RIS-aided links scales with the square of the number of REs~\cite{Liang2019, Wu2019Joint}. Therefore, the passive RIS would like to consume all the power budget to deploy the REs as many as possible. Consider the power budget $P_\mathrm{RIS}$, and the passive RIS can support $Q_{P} = \left\lfloor\frac{P_\mathrm{RIS}}{P_\mathrm{DC}}\right\rfloor$. This advantage becomes particularly significant in scenarios where the direct links among users are obstructed~\cite{Zhi2022ActiveRIS}. The performance of passive RIS in maximizing the sum rate in interference channels can be analyzed by addressing the following optimization problem
\begin{subequations}\label{prob:p1_pRIS}
  \begin{align}
    \max_{\boldsymbol{a}}&~~~ \sum^K_{k=1} R_{\mathrm{p},k}(\boldsymbol{a}) \label{eq:p1obj}\\
    \mathrm{s.t.}&~~~ 0\leq \alpha_q \leq 1, q = 1,\cdots, Q_{P}. \label{eq:PRIS_p1c2}
  \end{align}
\end{subequations}
This problem can be solved with the FP technique in Algorithm~\ref{alg:fp}. Notice that solving Problem~\eqref{prob:p1_pRIS} provides an upper-bound for the passive RIS-aided sum-rate maximization with the constraint~\eqref{eq:PRIS_p1c2}. For the passive RIS with the unit modulus constraint, its performance is investigated in~\cite{jiang2022interference}. For the RIS power minimization in interference channels, it cannot be directly investigated with the passive RIS design, because the power consumption of the passive RIS only depends on the number of REs and cannot be further optimized by the reflect beamforming design. To compare the passive RIS with the active RIS, an alternative feasible problem is considered in this paper that investigates the feasibility and the power consumption when the structure of the passive RIS and the number of REs are fixed. The considered feasible problem for the passive RIS is given by
\begin{subequations}\label{prob:p2_PRIS}
  \begin{align}
    \mathrm{find}&~~~ {\boldsymbol{a}} \label{eq:p2obj_PRIS}\\
    \mathrm{s.t.}&~~~R_{\mathrm{p},k}(\boldsymbol{a}) \geq {R}_{k}, \forall k \label{eq:p2c1_PRIS} \\
    &~~~ 0\leq \alpha_q \leq 1, q = 1,\cdots,Q. \label{eq:p2c2_PRIS}
  \end{align}
\end{subequations}

\section{Numerical Results}\label{sec:sim}

\begin{figure}[t]
  \centering
  \includegraphics[width=.99\columnwidth]{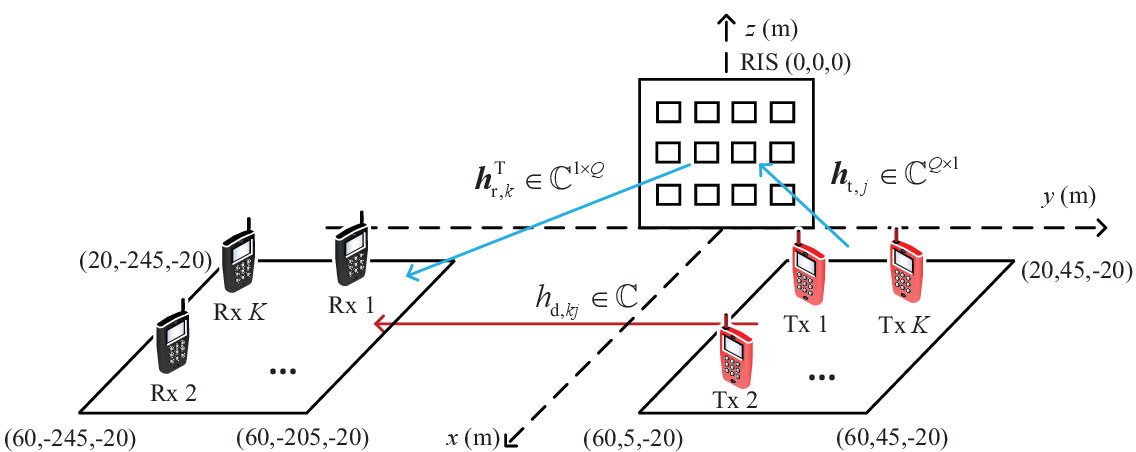}
  \caption{The simulated RIS-aided $K$ user pairs interference channels.}
  \label{fig:sim_model}
\end{figure}

In this section, numerical examples are provided to show the performance of the proposed sparse reflect beamforming designs for active RIS-aided interference channel systems. Without loss of generality, we consider the RIS-aided interference channels shown in Fig.~\ref{fig:sim_model}. The $K=4$ transmitters and their corresponding receivers are uniformly and randomly distributed in rectangular areas depicted by the $x$-$y$ coordinates $[20\mathrm{m},5\mathrm{m}]\times[60\mathrm{m},45\mathrm{m}]$ and $[20\mathrm{m},-245\mathrm{m}]\times[60\mathrm{m},-205\mathrm{m}]$, respectively, and the $z$-coordinates of all users are set to be $z=-20\mathrm{m}$. A $Q_1 \times Q_2$ uniform planar array-based RIS is deployed at the $y$-$z$ plane with $Q_1$ elements per row and $Q_2$ elements per column. The first RE of the RIS (the reference point) is located at $(0,0,0)$. The direct channel follows Rayleigh fading channel, and the forward and backward channels follow Rician fading. Consequently, the channels $\boldsymbol{h}_{i,k}$, where $i \in {\mathrm{t},\mathrm{r}}$, are modeled as follows
\begin{align}
  \boldsymbol{h}_{i,k} = \rho_{i,k}\left(\sqrt{\frac{\kappa}{1+\kappa}}\boldsymbol{h}_{i,k}^\mathrm{LOS}+\sqrt{\frac{1}{1+\kappa}}\boldsymbol{h}_{i,k}^\mathrm{NLOS}\right),
\end{align}
where $\rho^2_{i,k}$ is the corresponding pathloss from the RIS to Tx (Rx) $k$, $\kappa$ is the Rician factor with $\kappa=9$, the none-line-of-sight components are characterized by the complex Gaussian distribution $\mathcal{CN}(\boldsymbol{0},\mathbf{I}_Q)$, and the line-of-sight components are characterized by the RIS steering vectors $\boldsymbol{h}_{i,k}^\mathrm{LOS} = \boldsymbol{v}$. The $q$-th elements of the RIS steering vector is given by  
\begin{align}
  [\boldsymbol{v}(\theta_k,\varphi_k)]_q=e^{1j\frac{2\pi}{\lambda}\left(i_1(q)d_1\sin(\theta_k)\cos(\varphi_k)+i_2(q)d_2\sin(\varphi_k)\right)}, \nonumber
\end{align}
where $\theta_k$ and $\varphi_k$ are the corresponding azimuth and elevation angles between Tx (Rx) $k$ and the RIS, $\lambda$ is the wavelength of the signal carrier, $d_1=\lambda/2$ and $d_2=\lambda/2$ are the horizontal and vertical spacings between adjacent REs, respectively, and $i_1(q)=\mathrm{mod}(q-1,Q_1)$ and $i_2(q)=\left\lfloor (q-1)/Q_2\right\rfloor$ are the horizontal and vertical indices of element $q$, respectively. The noise power at the active RIS and at each of the $K$ receivers is assumed to be $\sigma_{r}^2=\sigma_s^2=-100\mathrm{dBm}$. The other simulation parameters are summarized in Table~\ref{table:sim}. All the averaging results are obtained with $500$ individual channel realizations. 

\begin{table}[!t]
  \caption{Simulation Parameters}\label{table:sim}
  \centering
  \begin{tabular}{|c|c|}  
  \hline
  Parameters & Values \\ \hline
  Pathloss for $\boldsymbol{h}_{\mathrm{t},k}$ and $\boldsymbol{h}_{\mathrm{r},k}$ (dB) & $-30-22\lg(d)$ \\ \hline
  Pathloss for ${h}_{\mathrm{d},kj}$ & $-30-40\lg(d)$ \\ \hline
  DC power $P_\mathrm{DC}$ & $-10\mathrm{dBm}$\\ \hline
  Biasing power $P_\mathrm{bias}$ & $-6\mathrm{dBm}$ \\ \hline
  \end{tabular}
\end{table}

\begin{figure}[t]
  \centering
  \includegraphics[width=.99\columnwidth]{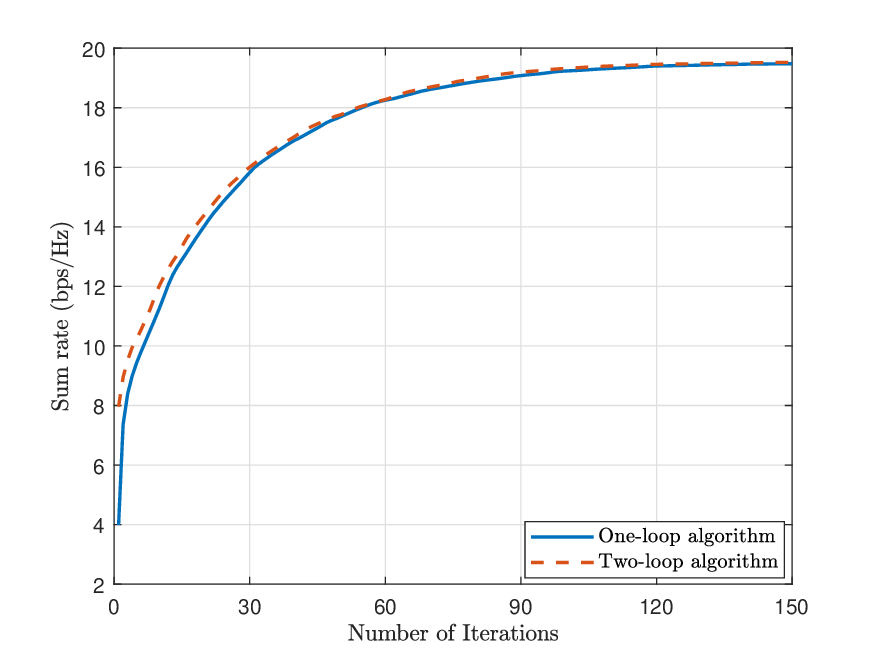}
  \caption{Convergence performance: $P_\mathrm{RIS}=10\mathrm{dBm}$ and $p_k=23\mathrm{dBm}$}
  \label{fig:sim_ite}
\end{figure}

\subsection{Sum-Rate Maximization}
For the sum-rate maximization problem, we first investigate the convergence performance of the proposed two-loop and one-loop algorithms. Specifically, we examine the number of outer-loop iterations required for the convergence of the two-loop algorithm. In this experiment, an $8\times 8$ RIS is considered. As shown in Fig.~\ref{fig:sim_ite}, solving the interference power minimization problem generates an initial point with a relatively small value. As the number of iterations increases, the sum-rate values for both algorithms rise monotonically, with a dramatic increase initially, followed by a gradual convergence to stationary values within approximately $110$ iterations. However, since the one-loop algorithm only solves the FP problem once for each iteration, it generally has a lower computational complexity than the two-loop algorithm. Thanks to the superiority of the one-loop algorithm in terms of computational complexity, the subsequent results related to sum-rate maximization are presented using this algorithm.

\begin{figure}[t]
  \centering
  \includegraphics[width=.99\columnwidth]{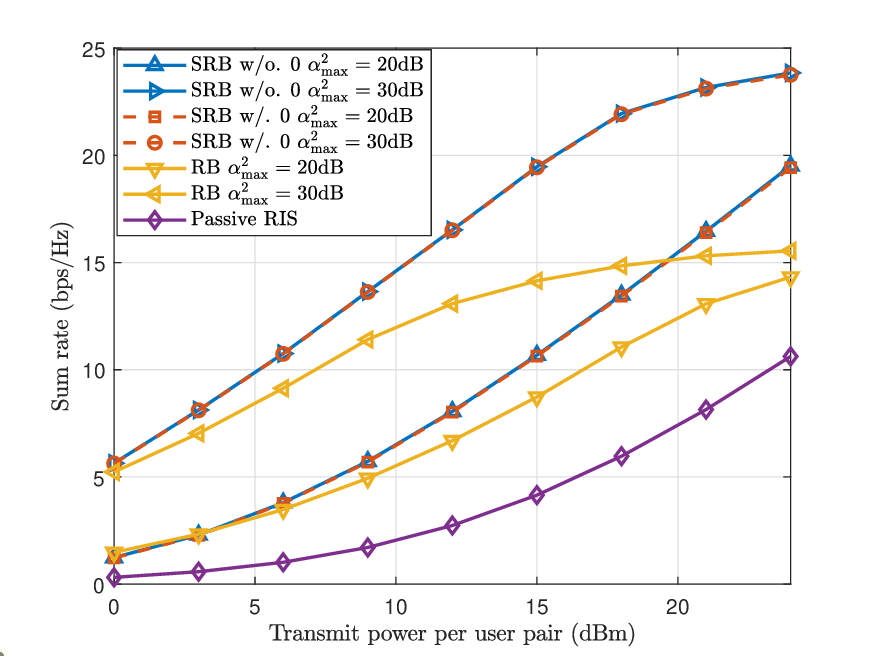}
  \caption{Transmit power versus sum rate: $P_\mathrm{RIS}=10\mathrm{dBm}$}
  \label{fig:sim_pk}
\end{figure}

As aforementioned in Sec.~\ref{sec:comparison}, when the power supply of the active RIS is sufficient, the active RIS tends to activate all the REs to amplify the incident signal for the purpose of maximizing the sum rate. To show the superiority of the proposed sparse reflect beamforming design for active RISs, we later compare the following four schemes under the insufficient power budget:
\begin{itemize}
  \item \emph{Sparse reflect beamforming (SRB) w/o. zero setting:} The sparse reflect beamforming vector $\boldsymbol{a}$ is obtained directly using Algorithm~\ref{alg:SRB_low}.
  \item \emph{SRB w/. zero setting:} This approach involves an additional zero-setting step on the vector generated by Algorithm~\ref{alg:SRB_low}. Specifically, elements with amplitudes less than unity are set to zero.
  \item \emph{Reflect beamforming (RB):} The reflect beamforming vector $\boldsymbol{a}$ is determined by addressing the sum-rate maximization problem as defined in~\eqref{prob:p1}. This approach is aligned with the fixed active RIS design strategy described in Sec.~\ref{sec:comparison}. Specifically, it involves activating the first $Q_{F}$ REs of the active RIS. The reflection coefficients for these activated REs are then calculated by solving the sum-rate maximization problem\eqref{prob:p1}, employing the original power consumption model as presented in~\eqref{eq:ori_power_model}.
  \item \emph{Passive RIS:} The passive RIS consumes the entire power budget to deploy a maximum number of REs, denoted by $Q_{P} = \left\lfloor{\frac{P_\mathrm{RIS}}{P_\mathrm{DC}}}\right\rfloor$. This corresponds to the passive RIS design outlined in Sec.\ref{sec:comparison}. The reflect beamforming vector for passive RIS is calculated by solving Problem\eqref{prob:p1_pRIS}.
\end{itemize}

As shown in Fig.~\ref{fig:sim_pk}, the proposed SRB schemes generally outperform the conventional RB scheme. This is because the proposed SRB scheme allows the active RIS to dynamically close the inefficient REs whose corresponding channel gains are not sufficiently high, and thus more power budget can be allocated to realize the high-quality amplification for these REs with high channel gains. However, for the conventional RB scheme, most of the power budget is used to support the biasing source power of $Q_F$ REs, and thus less power budget is left for the amplification. Moreover, it is also observed that in the high transmit power regime, the increases in the sum rate of the RB scheme with $\alpha_{\max}^2 = 30\mathrm{dB}$ is moderate. This is due to the fact that as the increase in the transmit power, more amplification power is needed according to~\eqref{eq:ori_power_model}, which the RB scheme with the limited power budget cannot afford. In addition, by comparing the sum rate between the SRB schemes with and without zero setting, the two schemes have  almost identical performance, indicating that Algorithm~\ref{alg:SRB_low} is capable of generating a high-quality solution with sparsity. Moreover, all the schemes related to the active RIS are superior to those of the passive RIS as expected. Though the passive RIS is able to acquire considerable performance gain by deploying more REs, the active RIS offers an effective alternative by directly amplifying the incident signal. Thanks to the requirements on less REs, the active RIS designs would be a preferable option for space-limited scenarios. 

\begin{figure}[t]
  \centering
  \includegraphics[width=.99\columnwidth]{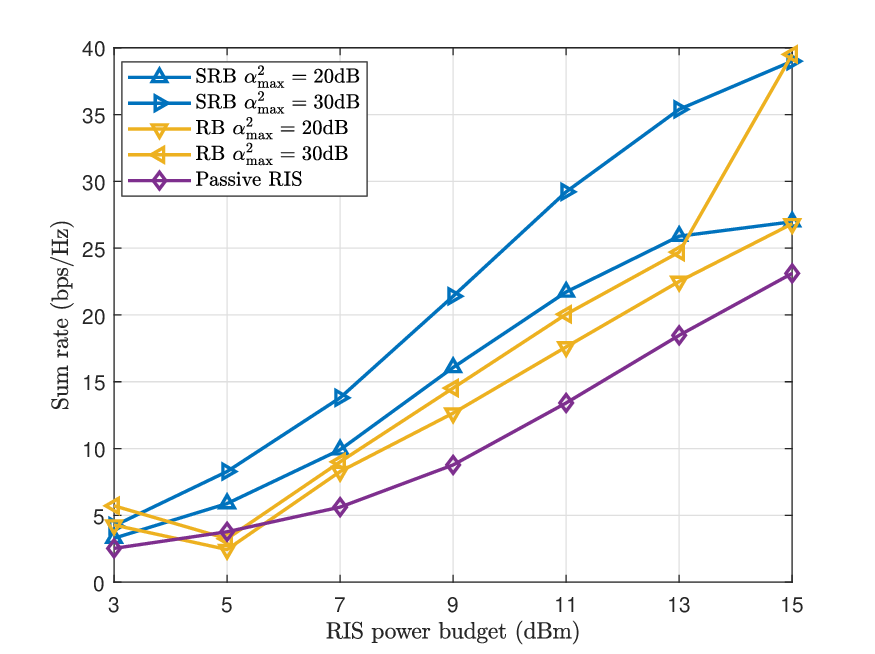}
  \caption{RIS power budget versus sum rate: $p_k=23\mathrm{dBm}$}
  \label{fig:sim_pris}
\end{figure}

Fig.~\ref{fig:sim_pris} is presented to investigate how the RIS power budget affects the sum rate performance of the RIS-aided interference channels. In this simulation, the SRB scheme is applied without zero setting. Notice that for the RB scheme, the curve is not smooth because of the discontinuity in the power model~\eqref{eq:ori_power_model} with respect to $Q_{F}$. For the SRB scheme, the curve is smooth because the power consumption model~\eqref{eq:power_modified} results in a smooth curve in the form of~\eqref{eq:power_budget_cst} after the $\ell_0$-norm relaxation. In general, the active RIS outperforms the passive RIS in the considered power budget regime except when the RB scheme is adopted with $P_\mathrm{RIS}=5\mathrm{dBm}$. This is because when $P_\mathrm{RIS}=5\mathrm{dBm}$, the most power budget is used to support $Q_{F}$ REs, and only a small amount power budget is left to support the amplification. As the RIS power budget grows up, the active RIS has moderate increases in the performance gains on the sum rate when compared with the passive RIS, and it seems that the passive RIS will eventually outperform the active RIS. This is because the active RIS herein has the limitation in the number of the REs ($Q_\mathrm{max} = 64$), and the passive RIS are allowed to consume the abundant power budget to deploy more REs without such a limitation. Then, let us pay attention to the comparisons between the SRB and the RB schemes. In most cases, the proposed SRB scheme works well with the limited power budget, while the considered RB scheme does not, as it suffers from the fixed RE selection. Besides, the proposed SRB scheme is applicable when there exists a sufficient power budget, which is shown by the observation that the SRB scheme has almost the same sum-rate performance of the RB scheme when $P_\mathrm{RIS} = 15\mathrm{dBm}$ ($Q_{F}=64$). Despite the performance gain, applying the SRB scheme gives us the opportunity to find a sparse reflect beamforming design with the affordable complexity, which avoids the need to solve the computationally expensive combinatorial problem.

\subsection{Active RIS Power Minimization}
For the active RIS power minimization problem, we first investigate the feasibility of meeting the rate requirement per user in interference channels. In this experiment, an $8\times 4$ RIS is considered, and the penalty factor of the DCA algorithm is set at $\rho = 10$. To further investigate the superiority of the proposed power-aware sparse reflect beamforming design, we compare the following three schemes
\begin{itemize}
  \item \emph{SRB:} The sparse reflect beamforming vector $\boldsymbol{a}$ is directly obtained by Algorithm~\ref{alg:SRB_apm}.
  \item \emph{RB:} The reflect beamforming vector $\boldsymbol{a}$ is obtained by solving the active RIS power minimization problem~\eqref{prob:p1} based on the fully active RIS design in Sec.~\ref{sec:comparison}, where all the REs are activated. The reflection coefficients of these REs are obtained by solving the active RIS power minimization problem with the power consumption model~\eqref{eq:ori_power_model}.
  \item \emph{Passive RIS:} The reflect beamforming vector $\boldsymbol{a}$ is obtained by solving the feasibility problem~\eqref{prob:p2_PRIS} with the passive RIS design. For the passive RIS with a given number of REs $Q_{P}$, the power consumption is calculated with~\eqref{eq:pRIS_power_model}, which corresponds to the passive RIS design in Sec.~\ref{sec:comparison}.
\end{itemize}

\begin{figure}[t]
  \centering
  \includegraphics[width=.99\columnwidth]{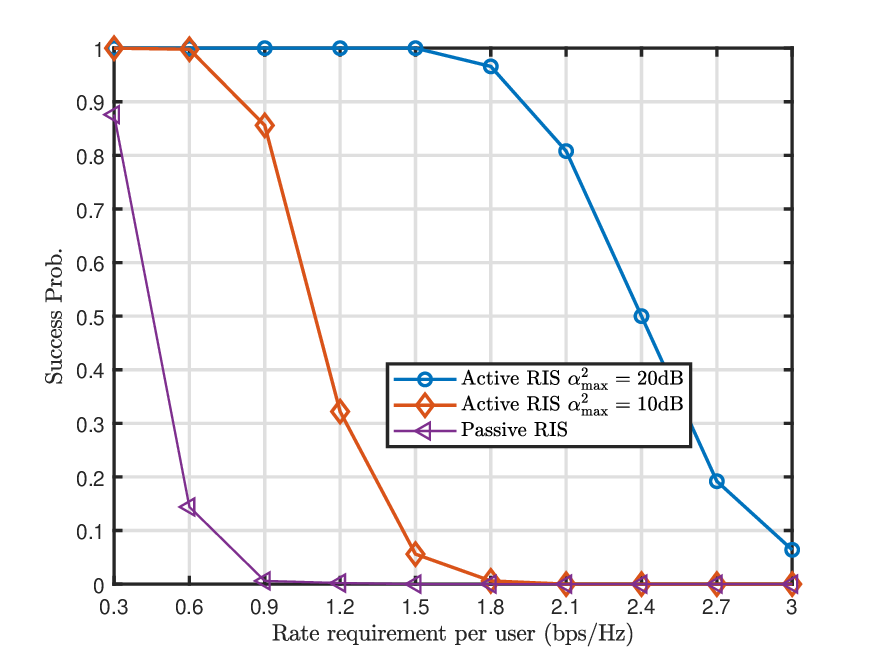}
  \caption{Rate requirement per user versus success probability: $p_k=23\mathrm{dBm}$}
  \label{fig:sim_success}
\end{figure}

\begin{figure}[t]
  \centering
  \includegraphics[width=.99\columnwidth]{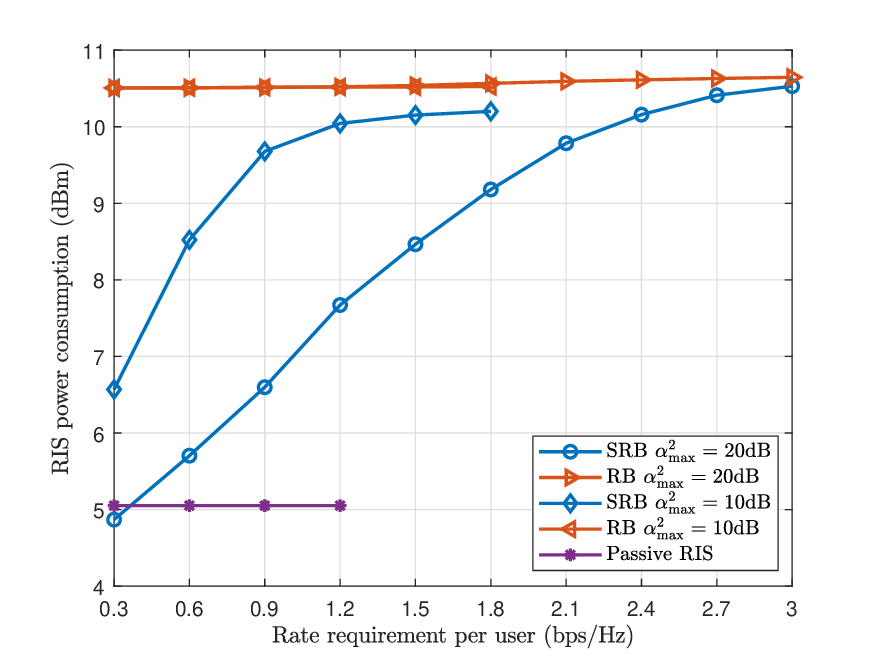}
  \caption{Rate requirement per user versus RIS power consumption: $p_k=23\mathrm{dBm}$}
  \label{fig:sim_minp}
\end{figure}

As shown in Fig.~\ref{fig:sim_success}, the success probabilities of various RIS optimization schemes are compared. It is important to note that both the proposed sparse reflect beamforming design and the conventional reflect beamforming design offer the same feasibility for this active RIS power minimization problem, as they share the same feasible set. Consequently, our primary comparison in Fig.~\ref{fig:sim_success} focuses on the success probabilities between active and passive RIS. In particular, the success probability of achieving rate requirements using passive RIS is determined by addressing Problem~\eqref{prob:p2_PRIS}. As expected, with the increase in the rate requirements, the success probabilities of all kinds of RIS decrease and eventually go to zero. Among them, the success probability of the passive RIS-aided system approaches zero when the rate requirement is $1.2~\mathrm{bps/Hz}$, while the active RIS with $\alpha_{\max}^2=10\mathrm{dB}$ can improve it to $1.8~\mathrm{bps/Hz}$. Increasing the amplification gain per RE, denoted by $\alpha_{\max}^2$, in an active RIS effectively boosts success probability, leading to improved capacity in interference channels.

Fig.~\ref{fig:sim_minp} shows the RIS power consumption by varying the rate requirements per user pair. Notice that the curves are plotted by averaging the results in which the RIS successfully supports the user pairs with the desired rates. The absence of some data points is due to the fact that the rate requirements cannot be fulfilled under the current schemes. In general, the passive RIS scheme consumes the least power under the given rate requirements, since it does not consume the biasing power to amplify the incident signals. However, it has the worst performance when it comes to the success probability in Fig.~\ref{fig:sim_success}. For the two schemes related to the use of active RIS, although the SRB and the RB schemes share the same success probability, the SRB scheme outperforms the RB scheme in terms of power consumption. With the same amplification gain, the RIS power consumption of the SRB scheme is gradually increasing with the rate requirements and finally converges to the curve of the RB scheme, but that of the RB scheme is slightly increasing. This occurs because the SRB scheme gradually activates more REs to meet increasingly stringent rate requirements, whereas the RB scheme keeps all REs constantly activated regardless of the stringency of these requirements. Hence, the RB scheme in fact provides an upper bound for the SRB scheme. The results also indicate that activating the REs consumes the most power, while directly amplifying signal requires relatively less power. Therefore, it is of great significance to study the sparse design on the active RIS in order to save power. 

In addition, by comparing the curves under different amplification gains for the SRB scheme, it is observed that improving the amplification gain facilitates energy saving of active RIS. An interesting observation of the SRB scheme occurs at a rate requirement of $0.3~\mathrm{bps/Hz}$ and an amplification gain of $\alpha_{\max}^2=20\mathrm{dB}$. This scenario demonstrates that active RIS can potentially consume less power to meet rate requirements with a higher success probability compared to passive RIS. This is somehow exciting, as the active RIS with the high amplification gain holds the promise to save power together with the sparse reflect beamforming design.

\section{Conclusion}\label{sec:con}
This paper has investigated active RIS-aided interference channels where $K$ user pairs transmit in the same time over a common frequency band with the assistance of active RIS. We have studied how the maximum amplitude constraint on RE affects the capability of RIS mitigating interference by solving the interference power minimization problem. Furthermore, we have considered the power-aware design for active RIS whose power consumption mainly depends on the number of activated REs. Based on this model, we have maximized the sum rate of the interference channel system subject to the maximum amplitude and the power budget constraints, and have also minimized the active RIS power consumption subject to the maximum amplitude and the minimum rate requirements. The sparse reflect beamforming vector solution to these problems has been obtained with the iterative $\ell_1$-norm reweighted algorithm. Numerical results have shown the superiority of the proposed power-aware designs for active RIS.

\bibliographystyle{IEEEtran}
\bibliography{library.bib}

\begin{thebibliography}{10}
\providecommand{\url}[1]{#1}
\csname url@samestyle\endcsname
\providecommand{\newblock}{\relax}
\providecommand{\bibinfo}[2]{#2}
\providecommand{\BIBentrySTDinterwordspacing}{\spaceskip=0pt\relax}
\providecommand{\BIBentryALTinterwordstretchfactor}{4}
\providecommand{\BIBentryALTinterwordspacing}{\spaceskip=\fontdimen2\font plus
\BIBentryALTinterwordstretchfactor\fontdimen3\font minus \fontdimen4\font\relax}
\providecommand{\BIBforeignlanguage}[2]{{%
\expandafter\ifx\csname l@#1\endcsname\relax
\typeout{** WARNING: IEEEtran.bst: No hyphenation pattern has been}%
\typeout{** loaded for the language `#1'. Using the pattern for}%
\typeout{** the default language instead.}%
\else
\language=\csname l@#1\endcsname
\fi
#2}}
\providecommand{\BIBdecl}{\relax}
\BIBdecl

\bibitem{you2021towards}
X.~You, C.-X. Wang, J.~Huang, X.~Gao, Z.~Zhang, M.~Wang, Y.~Huang, C.~Zhang, Y.~Jiang, J.~Wang \emph{et~al.}, ``Towards {6G} wireless communication networks: Vision, enabling technologies, and new paradigm shifts,'' \emph{Science China Information Sciences}, vol.~64, pp. 1--74, 2021.

\bibitem{Liang2019}
Y.-C. Liang, R.~Long, Q.~Zhang, J.~Chen, H.~V. Cheng, and H.~Guo, ``Large intelligent surface/antennas {LISA}: Making reflective radios smart,'' \emph{J. Commun. Inf. Netw.}, vol.~4, no.~2, pp. 40--50, Jun. 2019.

\bibitem{DiRenzo2020}
M.~{Di}~Renzo, A.~Zappone, M.~Debbah, M.-S. Alouini, C.~Yuen, J.~de~Rosny, and S.~Tretyakov, ``Smart radio environments empowered by reconfigurable intelligent surfaces: How it works, state of research, and the road ahead,'' \emph{IEEE J. Sel. Areas Commun.}, vol.~38, no.~11, pp. 2450--2525, 2020.

\bibitem{Guo2020Weighted}
H.~Guo, Y.-C. Liang, J.~Chen, and E.~G. Larsson, ``Weighted sum-rate maximization for reconfigurable intelligent surface aided wireless networks,'' \emph{IEEE Trans. Wireless Commun.}, vol.~19, no.~5, pp. 3064--3076, 2020.

\bibitem{Huang2019}
C.~{Huang}, A.~{Zappone}, G.~C. {Alexandropoulos}, M.~{Debbah}, and C.~{Yuen}, ``Reconfigurable intelligent surfaces for energy efficiency in wireless communication,'' \emph{IEEE Trans. Wireless Commun.}, vol.~18, no.~8, pp. 4157--4170, Aug 2019.

\bibitem{Long2021Active}
R.~Long, Y.-C. Liang, Y.~Pei, and E.~G. Larsson, ``Active reconfigurable intelligent surface-aided wireless communications,'' \emph{IEEE Trans. Wireless Commun.}, vol.~20, no.~8, pp. 4962--4975, 2021.

\bibitem{najafi2020physics}
M.~Najafi, V.~Jamali, R.~Schober, and H.~V. Poor, ``Physics-based modeling and scalable optimization of large intelligent reflecting surfaces,'' \emph{IEEE Trans. Commun.}, vol.~69, no.~4, pp. 2673--2691, 2020.

\bibitem{Wu2022Wideband}
L.~Wu, K.~Lou, J.~Ke, J.~Liang, Z.~Luo, J.~Y. Dai, Q.~Cheng, and T.~J. Cui, ``A wideband amplifying reconfigurable intelligent surface,'' \emph{IEEE Trans. Antennas Propag.}, vol.~70, no.~11, pp. 10\,623--10\,631, 2022.

\bibitem{Rao2023An}
J.~Rao, Y.~Zhang, S.~Tang, Z.~Li, C.-Y. Chiu, and R.~Murch, ``An active reconfigurable intelligent surface utilizing phase-reconfigurable reflection amplifiers,'' \emph{IEEE Trans. Microw. Theory Techn.}, vol.~71, no.~7, pp. 3189--3202, 2023.

\bibitem{Zhang2023Active}
Z.~Zhang, L.~Dai, X.~Chen, C.~Liu, F.~Yang, R.~Schober, and H.~V. Poor, ``Active {RIS} vs. passive {RIS}: Which will prevail in {6G}?'' \emph{IEEE Trans. Commun.}, vol.~71, no.~3, pp. 1707--1725, 2023.

\bibitem{Dong2022Active}
L.~Dong, H.-M. Wang, and J.~Bai, ``Active reconfigurable intelligent surface aided secure transmission,'' \emph{IEEE Trans. Veh. Technol.}, vol.~71, no.~2, pp. 2181--2186, 2022.

\bibitem{Gao2022Beamforming}
Y.~Gao, Q.~Wu, G.~Zhang, W.~Chen, D.~W.~K. Ng, and M.~{Di}~Renzo, ``Beamforming optimization for active intelligent reflecting surface-aided swipt,'' \emph{IEEE Trans. Wireless Commun.}, pp. 1--1, 2022.

\bibitem{Yang2023Active}
S.~Yang, R.~Long, and Y.-C. Liang, ``Active reconfigurable intelligent surface-aided cognitive radio system,'' in \emph{IEEE ICC}, 2023, pp. 1--6.

\bibitem{Jorswieck2008Complete}
E.~A. Jorswieck, E.~G. Larsson, and D.~Danev, ``Complete characterization of the pareto boundary for the {MISO} interference channel,'' \emph{IEEE Trans. Signal Process.}, vol.~56, no.~10, pp. 5292--5296, 2008.

\bibitem{cadambe2008interference}
V.~R. Cadambe and S.~A. Jafar, ``Interference alignment and degrees of freedom of the $ k $-user interference channel,'' \emph{IEEE Trans. Inf. Theory}, vol.~54, no.~8, pp. 3425--3441, 2008.

\bibitem{bafghi2022degrees}
A.~H.~A. Bafghi, V.~Jamali, M.~Nasiri-Kenari, and R.~Schober, ``Degrees of freedom of the {K}-user interference channel assisted by active and passive {IRSs},'' \emph{IEEE Trans. Commun.}, 2022.

\bibitem{jiang2022interference}
T.~Jiang and W.~Yu, ``Interference nulling using reconfigurable intelligent surface,'' \emph{IEEE J. Sel. Areas Commun.}, vol.~40, no.~5, pp. 1392--1406, 2022.

\bibitem{elmossallamy2021ris}
M.~A. ElMossallamy, K.~G. Seddik, W.~Chen, L.~Wang, G.~Y. Li, and Z.~Han, ``{RIS} optimization on the complex circle manifold for interference mitigation in interference channels,'' \emph{IEEE Trans. Veh. Technol.}, vol.~70, no.~6, pp. 6184--6189, 2021.

\bibitem{Amato2018a}
F.~{Amato}, H.~M. {Torun}, and G.~D. {Durgin}, ``{RFID} backscattering in long-range scenarios,'' \emph{IEEE Trans. Wireless Commun.}, vol.~17, no.~4, pp. 2718--2725, April 2018.

\bibitem{chen2019channel}
J.~Chen, Y.-C. Liang, H.~V. Cheng, and W.~Yu, ``Channel estimation for reconfigurable intelligent surface aided multi-user {MIMO} systems,'' \emph{arXiv preprint arXiv:1912.03619}, 2019.

\bibitem{candes2008enhancing}
E.~J. Candes, M.~B. Wakin, and S.~P. Boyd, ``Enhancing sparsity by reweighted $\ell_1$ minimization,'' \emph{J. Fourier Anal. Appl.}, vol.~14, no.~5, pp. 877--905, 2008.

\bibitem{Dai2014Sparse}
B.~Dai and W.~Yu, ``Sparse beamforming and user-centric clustering for downlink cloud radio access network,'' \emph{IEEE Access}, vol.~2, pp. 1326--1339, 2014.

\bibitem{Shen2018Fractional}
K.~Shen and W.~Yu, ``Fractional programming for communication systems—part {I}: Power control and beamforming,'' \emph{IEEE Trans. Signal Process.}, vol.~66, no.~10, pp. 2616--2630, 2018.

\bibitem{grant2008cvx}
\BIBentryALTinterwordspacing
M.~Grant and S.~Boyd, ``{CVX}: Matlab software for disciplined convex programming,'' 2008. [Online]. Available: \url{cvxr.com/cvx.}
\BIBentrySTDinterwordspacing

\bibitem{lobo1998applications}
M.~S. Lobo, L.~Vandenberghe, S.~Boyd, and H.~Lebret, ``Applications of second-order cone programming,'' \emph{Linear algebra and its applications}, vol. 284, no. 1-3, pp. 193--228, 1998.

\bibitem{Wu2019}
Q.~{Wu} and R.~{Zhang}, ``Intelligent reflecting surface enhanced wireless network via joint active and passive beamforming,'' \emph{IEEE Trans. Wireless Commun.}, vol.~18, no.~11, pp. 5394--5409, Nov 2019.

\bibitem{Yang2020Federated}
K.~Yang, T.~Jiang, Y.~Shi, and Z.~Ding, ``Federated learning via over-the-air computation,'' \emph{IEEE Trans. Wireless Commun.}, vol.~19, no.~3, pp. 2022--2035, 2020.

\bibitem{le2018dc}
H.~A. Le~Thi and T.~Pham~Dinh, ``{DC} programming and {DCA}: thirty years of developments,'' \emph{Mathematical Programming}, vol. 169, no.~1, pp. 5--68, 2018.

\bibitem{rockafellar2015convex}
R.~T. Rockafellar, ``Convex analysis,'' in \emph{Convex Analysis}.\hskip 1em plus 0.5em minus 0.4em\relax Princeton University Press, 2015.

\bibitem{Luo2010}
Z.-Q. Luo, W.-K. Ma, A.~M.-C. So, Y.~Ye, and S.~Zhang, ``Semidefinite relaxation of quadratic optimization problems,'' \emph{IEEE Signal Processing Magazine}, vol.~27, no.~3, pp. 20--34, 2010.

\bibitem{Wu2019Joint}
Q.~Wu and R.~Zhang, ``Joint active and passive beamforming optimization for intelligent reflecting surface assisted {SWIPT} under {QoS} constraints,'' \emph{arXiv preprint arXiv:1910.06220}, 2019.

\bibitem{Zhi2022ActiveRIS}
K.~Zhi, C.~Pan, H.~Ren, K.~K. Chai, and M.~Elkashlan, ``Active {RIS} versus passive {RIS}: Which is superior with the same power budget?'' \emph{IEEE Commun. Lett.}, vol.~26, no.~5, pp. 1150--1154, 2022.

\end{thebibliography}
\end{document}